\documentclass[useAMS,usenatbib]{mnras}

\usepackage[dvips,dvipdfmx]{graphicx}

\usepackage{amsmath}
\usepackage[usenames,dvipsnames]{color}

\def\lesssim{\mathrel{\hbox{\rlap{\hbox{\lower5pt\hbox{$\sim$}}}\hbox{$<$}}}}
\def\gtrsim{\mathrel{\hbox{\rlap{\hbox{\lower5pt\hbox{$\sim$}}}\hbox{$>$}}}}

\title[Parametric initial conditions for SN simulations] {Parametric
  initial conditions for core-collapse supernova simulations}
\author[Suwa and M\"uller]{ Yudai Suwa$^{1,2}$\thanks{E-mail:
    suwa@yukawa.kyoto-u.ac.jp} and Ewald M\"uller$^{2}$
  \\ $^{1}$Yukawa Institute for Theoretical Physics, Kyoto University,
  Oiwake-cho, Kitashirakawa, Sakyo-ku, Kyoto, 606-8502,
  Japan\\ $^{2}$Max-Planck-Institut f\"ur Astrophysik,
  Karl-Schwarzschild-Str. 1, D-85748 Garching, Germany }

\begin{document}

\date{Accepted. Received.}

\pagerange{\pageref{firstpage}--\pageref{lastpage}} \pubyear{2016}

\maketitle

\label{firstpage}

\begin{abstract}
  We investigate a method to construct parametrized progenitor models
  for core-collapse supernova simulations.  Different from all modern
  core-collapse supernova studies, which rely on progenitor models
  from stellar evolution calculations, we follow the methodology of
  \cite{baro90} to construct initial models. Choosing parametrized
  spatial distributions of entropy and electron fraction as a function
  of mass coordinate and solving the equation of hydrostatic
  equilibrium, we obtain the initial density structures of our
  progenitor models. First, we calculate structures with parameters
  fitting broadly the evolutionary model s11.2 of \cite{woos02}.  We
  then demonstrate the reliability of our method by performing general
  relativistic hydrodynamic simulations in spherical symmetry with the
  isotropic diffusion source approximation to solve the neutrino
  transport. Our comprehensive parameter study shows that initial
  models with a small central entropy ($\lesssim 0.4\,k_B$
  nucleon$^{-1}$) can explode even in spherically symmetric
  simulations. Models with a large entropy ($\gtrsim 6\,k_B$
  nucleon$^{-1}$) in the Si/O layer have a rather large explosion
  energy ($\sim 4\times 10^{50}$\,erg) at the end of the simulations,
  which is still rapidly increasing.
\end{abstract}


\section{Introduction}
\label{sec:intro}

Numerical simulations of core-collapse supernovae have been showing
significant rapid development during the last decade \citep[see][for
  recent reviews and references therein]{jank12, kota12, burr13,
  fogl15}. Especially, multi-dimensional neutrino-radiation
hydrodynamic simulation, which consistently solve the equations of
hydrodynamics and neutrino transport, become a doable problem thanks
to the advance in supercomputer performance and the development of
efficient numerical schemes. After \cite{bura06}, an increasing number
of authors have presented multi-dimensional simulations including some
treatment of spectral neutrino transport, which obtained explosions
(\citealt{mare09,suwa10,muel12b,brue13,pan16,ocon15} for 2D and
\citealt{taki12,mels15a,lent15,muel15b} for 3D), while other
simulations produced failures, i.e., no explosion by neutrino heating
\citep{burr06, ott08, hank13, dole15}.  Moreover, most of the
successful studies suffer from the {\it explosion energy problem},
that is, the explosion energies obtained in these simulations ($\sim
10^{50}$ erg) are much smaller than the canonical observed values
($\sim 10^{51}$ erg).

A supernova simulation is an initial value problem. In particular, the
initial conditions of such a simulation are based on stellar
evolutionary calculations. In these one-dimensional calculations many
assumptions and approximations are employed, especially to treat
multi-dimensional flows, such as e.g. convection, because currently
only one-dimensional (spherically symmetric) calculations are
feasible. However, multi-dimensional effects are supposed to make
explosions easier, as e.g. the non-radial velocity field inside
burning shells \citep{meak07, arne11, couc13c, muel15a, couc15b}.
Several authors have investigated the influence of the progenitor
properties systematically \citep{ugli12, ocon13, naka15, suwa16a,
  pejc15, ertl16, sukh15} using different progenitor models
\citep[especially from][]{woos02,limo06,woos07}. These studies have
revealed that the explosion characteristics strongly depend on the
mass of the progenitor and on its internal structure.  However, it is
still unclear which are the most important quantities among those
characterizing the internal structure of a core-collapse supernova
progenitor.

Because stellar evolutionary calculations are subject to restrictions
\citep[see, e.g.][for recent code comparison]{jone15}, we decided to
generate progenitor models by ourselves in a more systematic and
manageable way. To this end, we used the approach proposed by
\cite{baro90} to construct initial models. In this approach one
prescribes the distributions of entropy and electron fraction ($Y_e$)
in a progenitor model as functions of the mass coordinate, and one
assumes hydrostatic equilibrium to obtain the density structure from
these distributions. The hydrodynamic evolution of the progenitor
models is then simulated employing a microscopic equation of state.
We note that the temperature distribution is more crucial when one
wants to obtain initial modes whose energy generation rates are
consistent with those of stellar evolution models. However, since the
entropy distribution is better suited for characterizing parametrized
initial models, we use it instead of the temperature distribution in
this work.

Contrary to \cite{baro90}, we apply their approach to modern radiation
hydrodynamic simulations of neutrino-driven core-collapse supernovae.
While a neutrino-driven explosion is the current standard paradigm for
core-collapse supernovae, \cite{baro90} were discussing the influence
of progenitor properties on the prompt explosion scenario, in which
the prompt shock resulting from core bounce was thought to cause the
explosion.  In particular, we have performed one-dimensional (1D)
general relativistic hydrodynamic simulations including a detailed
treatment of neutrino transport and a nuclear equation of state, i.e.,
our study is more elaborate than that of \cite{baro90}.

Using this approach, we were able to perform a comprehensive parameter
study which displays the dependencies of the outcome of 1D
core-collapse supernova simulations on the properties of the
progenitor models.  
In addition, our approach has the advantage over other numerical
studies of core-collapse supernovae, which all rely on progenitor
models from stellar evolutionary calculations \citep[but
  see][]{yama16}, that broader initial conditions can be studied than
those currently predicted by stellar evolutionary calculations.

In section\,\ref{sec:progenitor} we explain how we constructed the
progenitor models, and in section\,\ref{sec:hydrodynamic} we describe
our hydrodynamic method and present the results of our simulations. We
discuss in detail the influence of the progenitor properties on the
core-collapse supernova dynamics in section\,\ref{sec:bc90}, and
conclude in section\,\ref{sec:summary} with a summary and discussion
of our results.

\section{Progenitor models}
\label{sec:progenitor}

In this section, we explain the strategies to obtain progenitor models
for core-collapse supernova simulations. First of all, we construct
progenitor models resembling the stellar evolutionary model s11.2 of
\cite{woos02}, which has been widely used in hydrodynamic simulations.

\subsection{Hydrostatic equation}

To construct a progenitor model for our hydrodynamic simulations, we
solve the hydrostatic equation
\begin{equation}
  \frac{dP}{dM} = -\frac{GM}{4\pi r^4},
\label{eq:hydrostatic}
\end{equation}
where $P, M, G$, and $r$ are the pressure, the mass coordinate, the
gravitational constant, and the radial coordinate, respectively. The
density is given by $dM/dr = 4\pi r^2\rho$.  To solve
Eq.\,(\ref{eq:hydrostatic}) one needs to specify a value for the
central density, $\rho_0$, which is one of the parameters of this
approach, and one needs to have $P$ given as a function of density
$\rho$, entropy $S$, and electron fraction $Y_e$, i.e., an equation of
state (EOS).

Following \cite{baro90}, we change $G \to g_\mathrm{eff} G$ in
Eq.\,(\ref{eq:hydrostatic}), where $g_\mathrm{eff}<1$ is a factor
mimicking the fact that the progenitors are no longer in hydrostatic
equilibrium, but already in a dynamic state. We used this procedure to
destabilize the core in a uniform way, because reducing instead the
pressure (by reducing the entropy or $Y_e$) may lead to undesirable
effects, like e.g., a strange mass accretion history
\cite[see][]{baro90}.

In the following subsections, we give the distributions of $S$ and
$Y_e$ as functions of the mass coordinate $M$ that we used in our
study. Given these functions, we integrate Eq.\,(\ref{eq:hydrostatic})
and obtain $\rho(r)$ and $M(r)$. Due to limited extent of the tabular
equation of state used in our simulations, we integrate
Eq.\,(\ref{eq:hydrostatic}) outward in mass until the density drops
below a value of $10^3$\,g\,cm$^{-3}$. We note that this Newtonian
treatment of the progenitor model is compatible with the general
relativistic treatment used in our hydrodynamic simulations, because
the central lapse function is $1 - \alpha\approx O(10^{-3})$ for the
progenitor models, i.e., the use of the Newtonian approximation is
well justified.

\subsection{Entropy and electron fraction}

\begin{figure}
\centering
\includegraphics[width=0.45\textwidth]{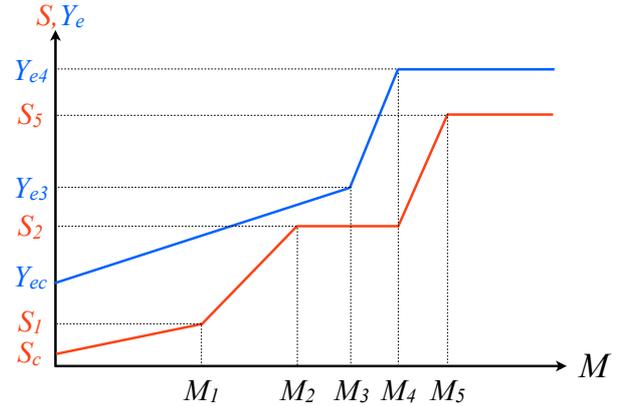}
\caption{ Schematic behavior of the entropy $S$ (red line) and
  electron fraction $Y_e$ (blue line) distribution as a function of
  mass for our progenitor models. }
\label{fig:schematic}
\end{figure}

Following \cite{baro90} the entropy as a function of mass $M$ of our
progenitor models is given by
\begin{align}
S(M\le M_1)       &= S_c + (S_1-S_c) \left(\frac{M}{M_1}        \right),\\
S(M_1\le M\le M_2)&= S_1 + (S_2-S_1) \left(\frac{M-M_1}{M_2-M_1}\right),\\
S(M_2\le M\le M_4)&= S_2 = S_3 = S_4                                   ,\\
S(M_4\le M\le M_5)&= S_4 + (S_5-S_4) \left(\frac{M-M_4}{M_5-M_4}\right),\\
S(M_5\le M)       &= S_5,
\end{align}
and the electron fraction by
\begin{align}
Y_e(M\le M_3)       & =Y_{ec} + (Y_{e3}-Y_{ec})
                                \left(\frac{M}{M_3}\right),\\
Y_e(M_3\le M\le M_4)&= Y_{e3} + (Y_{e4}-Y_{e3})
                                \left(\frac{M-M_c}{M_4-M_3}\right),\\
Y_e(M_4\le M)       &= Y_{e4},
\end{align}
(see also Fig.\,\ref{fig:schematic}). These two distributions contain
the parameters $M_1$, $M_2$, $M_3$, $M_4$, $M_5$, $S_c$, $S_1$, $S_2$,
$S_5$, $Y_{ec}$, $Y_{e3}$, and $Y_{e4}$, which uniquely characterize
the progenitor model.

The mass parameters have the following significance for an iron core
which is undergoing silicon shell-burning after the central convective
core has been exhausted of fuel. $M_1$ is the mass coordinate of the
edge of the final convection in the radiative core, $M_2$ is the mass
coordinate of the inner edge of the convection zone in the iron core,
$M_3$ is the mass coordinate up to which the core matter is in nuclear
statistical equilibrium (NSE core), $M_4$ is the iron core mass, and
$M_5$ is the mass coordinate at the base of the silicon/oxygen shell
which has a much larger entropy than the iron core ($S_5 \gg S_2$) and
consequently a much lower density.

Since the entropy distribution is a consequence of a complicated
sequence of burning and convection stages, its profile is more
structured than the $Y_e$ profile, because the latter only depends on
the electron capture timescale.  Different from \citet{baro90}, we
used two additional parameters, $M_5$ and $S_5$, in our study, because
in more recent stellar evolutionary models the locations do not
coincide where $S$ and $Y_e$ increase strongly. Also note that in
\citet{baro90} the iron core mass is given by $(M_3+M_4)/2$, whereas
it is $M_4$ in our study.

According to stellar evolutionary calculations, these above parameters
vary from progenitor model to progenitor model. For example, $S_c$
varies from $\sim 0.5$ to $\sim 2$, and the values of $M_5$ range from
$\sim 1.2$ to $\sim 3.7$ (see Table \ref{tab4} in the Appendix).

\begin{table*}
\centering
\caption{Parameters of our models resembling the stellar evolutionary model 
         s11.2 of \citet{woos02}}
\begin{tabular}{lccccccccccccccc}
\hline
Model & $M_1$ & $M_2$ & $M_3$ & $M_4$ & $M_5$ & $S_c$ & $S_1$ & $S_2$ & $S_5$ 
      & $Y_{ec}$ & $Y_{e3}$ & $Y_{e4}$ & $\rho_c$ & $g_\mathrm{eff}$ \\
      & \multicolumn{5}{c|}{[$M_\odot$]} & \multicolumn{4}{c|}{[$k_B/$baryon]} 
      & & & & [$10^{10}$g cm$^{-3}$] \\
\hline
WHW02-s11.2-g0.99   & 0.82 & 1.16 & 1.26 & 1.30 & 1.32 & 0.62 & 1.1  & 1.74  & 5.4 
                    & 0.425 & 0.48 & 0.5 & 1.6 & 0.99\\ 
WHW02-s11.2-g0.975  & ---  & ---  & ---  & ---  & ---  & ---  & 1.0  & 1.65  & --- 
                    & --- & --- & --- & --- & 0.975\\ 
WHW02-s11.2-g0.95   & ---  & ---  & ---  & ---  & ---  & ---  & 0.75 & 1.64  & --- 
                    & --- & --- & --- & --- & 0.95\\ 
\hline\hline
\end{tabular}
\label{tab}
\end{table*}

\begin{figure}
\centering
\includegraphics[width=0.45\textwidth]{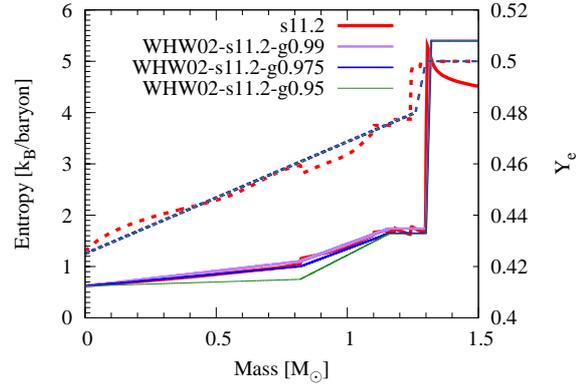}
\caption{Entropy (solid) and electron fraction (dotted) distribution
  as a function of mass coordinate. Red lines give the distributions
  of the stellar evolutionary model s11.2 of \citet{woos02}, and the
  purple, blue, and green lines show the distributions of our models,
  whose parameters are given in Table \ref{tab}.  }
\label{fig:s-ye_s112}
\end{figure}

Fig.\,\ref{fig:s-ye_s112} shows the $S$ (solid lines) and $Y_e$
(dotted lines) distributions of the three models investigated in this
study (purple, blue, and green lines) compared to those of the stellar
evolutionary model s11.2 (red line) of \cite{woos02}, which has been
used in many core-collapse supernova studies during the past decade.
In Table \ref{tab} we give the corresponding parameters of our three
progenitor models, which have the same $Y_e$ profile, but differ by
the values of $S_1$, $S_2$, and $g_\mathrm{eff}$.  We used different
values of $S_1$ and $S_2$ to match the density structure of the
stellar evolutionary model s11.2, and varied the value of
$g_\mathrm{eff}$ from 0.99 to 0.95.  As shown in
Fig.\,\ref{fig:density_s112}, all three models reproduce the density
structure of model s11.2 very well.

\subsection{Density structures}

\begin{figure}
\centering
\includegraphics[width=0.45\textwidth]{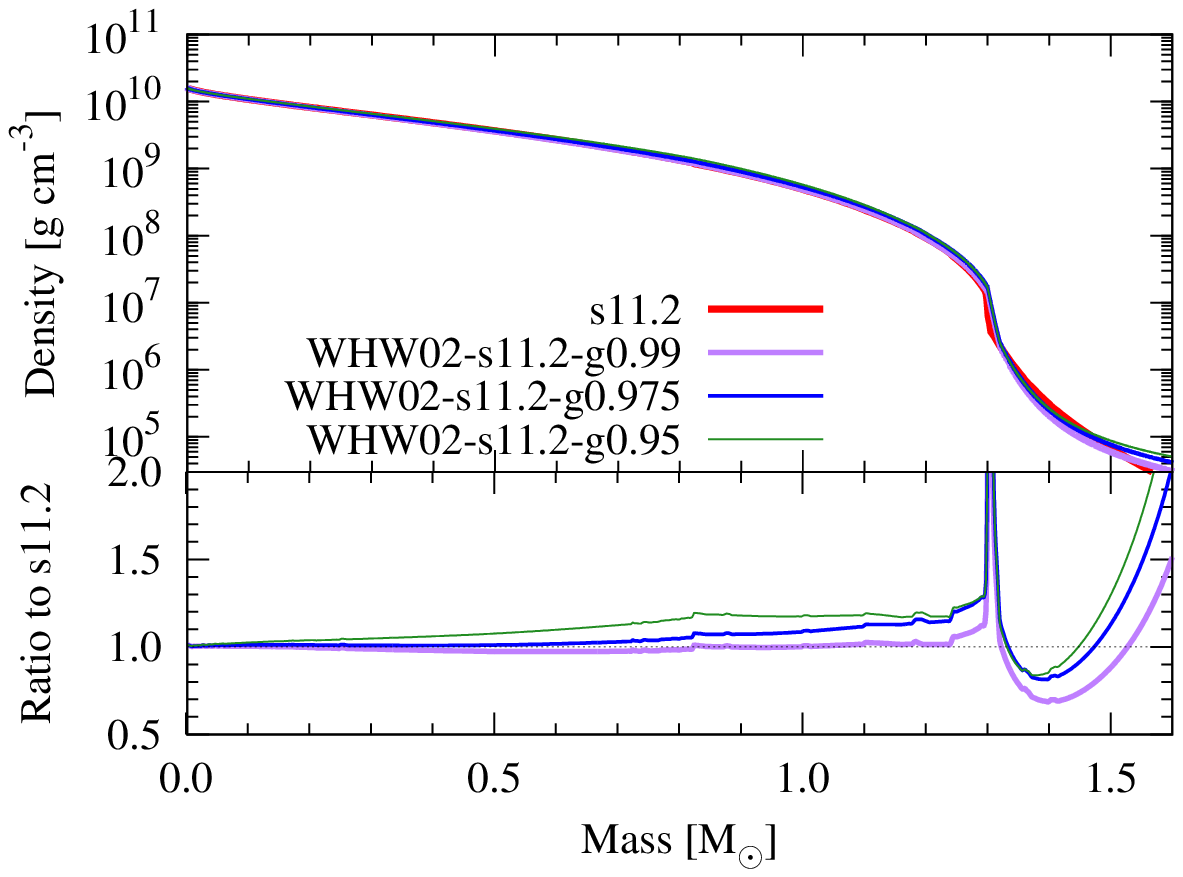}
\includegraphics[width=0.45\textwidth]{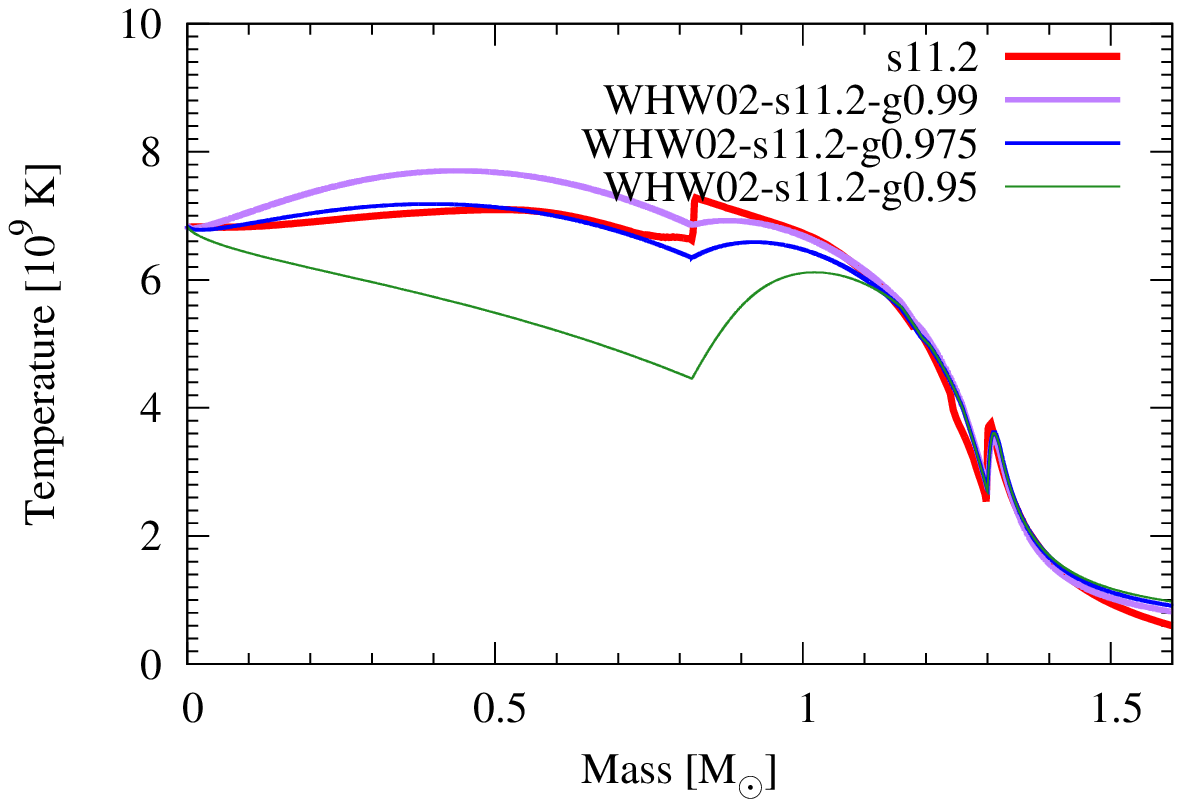}
\caption{Density (top panel) and temperature (bottom panel)
  distributions of the stellar evolutionary model s11.2 (red) and of
  our three corresponding progenitor models (purple, blue, and
  green). The bottom plot in the upper panel shows the density
  distributions of our three models normalized by that of the s11.2
  model. }
\label{fig:density_s112}
\end{figure}

The top panel of Fig.\,\ref{fig:density_s112} shows the density
structures of the stellar evolutionary model s11.2 together with those
of our three corresponding progenitor models, which are obtained by
integrating Eq.\,(\ref{eq:hydrostatic}) numerically. The bottom plot
of this panel, which displays the density distributions of our three
models normalized by that of the s11.2 model, proves that our three
models have almost the same density structure as model s11.2, the
error being less than a few 10\% except for the outermost layers ($M
\gtrsim 1.3M_\odot$) of our models.  On the other hand, the
temperature structures (bottom panel in Fig.\,\ref{fig:density_s112})
of all of our three models differ significantly from that of model
s11.2 for $M \lesssim 1.0M_\odot$ because of they have quite different
entropy profiles from the s11.2 model.

Obviously the parameterized progenitor models of \cite{baro90} can
reproduce the progenitor structure of stellar evolutionary models at
collapse considerably well. The model parameters that fit the
structural properties of other progenitor models are given in Appendix
\ref{sec:other}.

\section{Hydrodynamic simulations}
\label{sec:hydrodynamic}

\subsection{Method}
\label{sec:method}

For our simulations we used the code Agile-IDSA \citep{lieb09}, which
is a publicly available
\footnote{https://physik.unibas.ch/\~{}liebend/download/}
1D neutrino-radiation hydrodynamics code for simulating core-collapse
supernovae.  The hydro solver, Agile (Adaptive Grid with Implicit Leap
Extrapolation), integrates the general relativistic hydrodynamic
equations in spherical symmetry \citep{lieb02}, while the radiation
transport part is an implementation of the isotropic diffusion source
approximation (IDSA) \citep{lieb09}, which has been used, e.g., by
\cite{suwa10, taki12, naka15}, and \cite{pan16} to perform
multi-dimensional core-collapse simulations.  In IDSA the electron
neutrino and electron anti-neutrino distribution functions are split
into two components, which are solved with different numerical
techniques.

The weak interaction rates implemented in our code are based on
\cite{mezz93}, and the cooling by muon and tau neutrinos is modeled
with a leakage scheme.  Neutrino-electron scattering is also
implemented in this code according to \cite{lieb05b} by expressing the
electron fraction $Y_e$ as a function of $\rho$. However, since this
function is calibrated for specific progenitor models and it is not
always adequate, we did not employ it in this work.
The equation of state (EOS) used in our simulations is that of
\cite{latt91} with an incompressibility $K=220$\,MeV for $\rho \ge
10^8$\,g\,cm$^{-3}$ and that of \cite{timm99} for $\rho <
10^8$\,g\,cm$^{-3}$. In the latter density range the average nuclear
mass number $A$ and atomic number $Z$ are assumed to be the same as in
the EOS of \cite{latt91} at $\rho = 10^8$\,g\,cm$^{-3}$.  We follow
\cite{ocon10} to match the thermodynamic quantities of both EOS tables
at the transition density. The minimum density of of our combined EOS
table is $10^{3}$\,g\,cm$^{-3}$.

Accordingly, the results of our study are based on the use of a modern
numerical tool that is well suited for simulations of neutrino-driven
supernova explosions, because it is able to handle general
relativistic gravity, neutrino radiation transport, and a nuclear
equation of state. Nowadays we know that all of these ingredients are
of considerable importance for a proper simulation of the supernova
explosion mechanism, but none of them were taken into account in the
work of \cite{baro90}.

\subsection{Results}
\label{sec:result}

\begin{figure}
\centering
\includegraphics[width=0.45\textwidth]{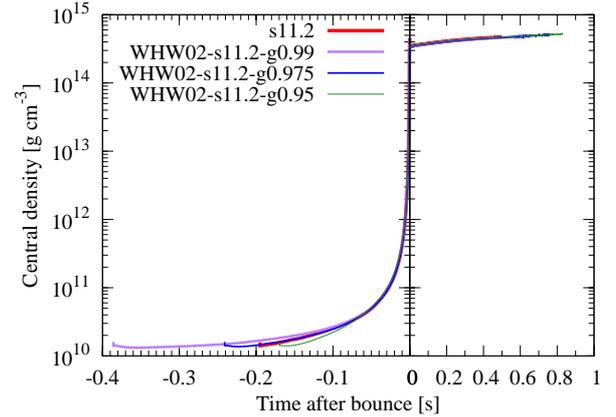}
\caption{Central density as a function of time after bounce for the
  stellar evolutionary model s11.2 (red) and our three corresponding
  progenitor models (purple, blue, and green). The time of bounce
  differs, being later for models with a larger $g_\mathrm{eff}$
  because these models are closer to a hydrostatic configuration.}
\label{fig:density_in_time}
\end{figure}

Fig.\,\ref{fig:density_in_time} shows the central density as a
function of time after bounce for all investigated models based on
s11.2. Because our models were computed with different values of
$g_\mathrm{eff}$, they bounce at different times, which range from
$\sim 390$ to 170\,ms.  The density evolution of the stellar
evolutionary model s11.2 is very similar to that of model
WHW02-s11.2-g0.975 (although the central density of the model slightly
decreases because the grid resolutions of the hydrodynamical
simulations and those of the initial models differ).  The figure
implies that the collapse of our initially hydrostatic models with
$g_\mathrm{eff} \lesssim 0.975$ proceeds similarly to that of the
already dynamically collapsing core of the stellar evolutionary
progenitor model s11.2, even though the former models do not have any
initial radial velocity.

\begin{figure}
\centering
\includegraphics[width=0.49\textwidth]{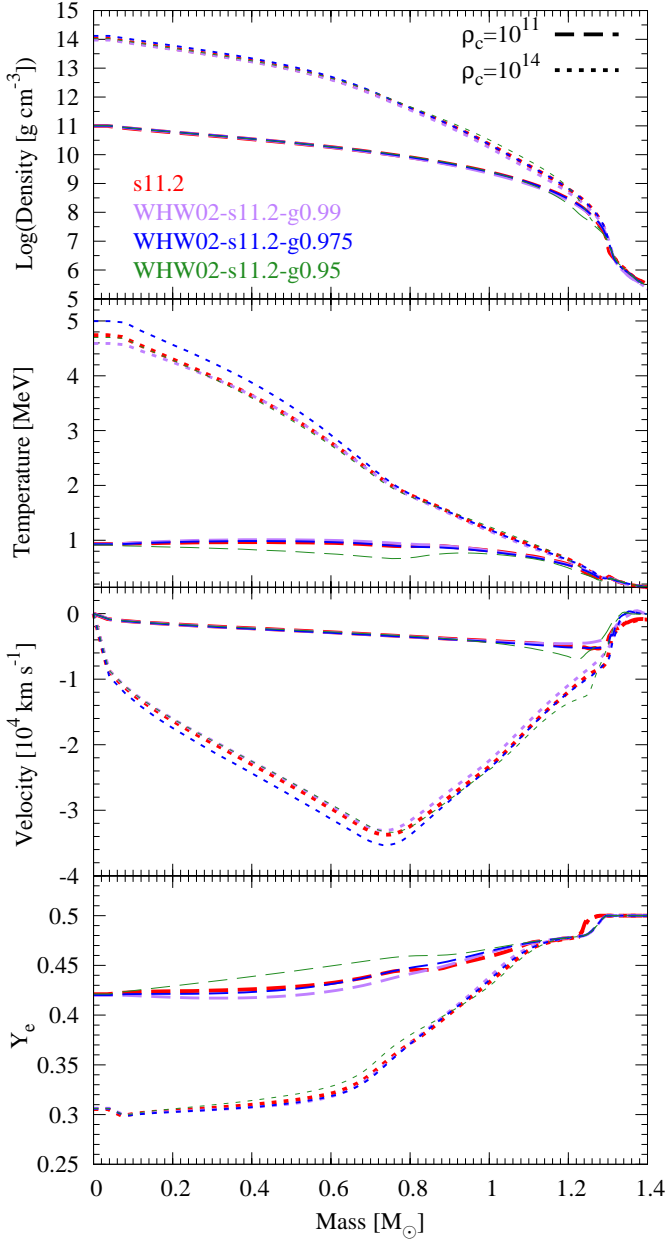}
\caption{Comparison of the density, temperature, radial velocity, and
  electron fraction distributions (from top to bottom).  Dashed and
  dotted lines give the profiles at the time when the central density
  reaches a value of about $10^{11}$ and $10^{14}$\,g\,cm$^{-3}$,
  respectively.}
\label{fig:s11.2}
\end{figure}

Fig.\,\ref{fig:s11.2} shows the evolution of the density, temperature,
radial velocity, and electron fraction distributions before core
bounce. The snapshots are taken at the time when the central density
has a value of $\approx 10^{11}$ (dashed lines) and $\approx
10^{14}$\,g\,cm$^{-3}$ (dotted lines), respectively. At the earlier
snapshot ($\rho_c = 10^{11}$\,g\,cm$^{-3}$), the temperature
distribution of model WHW02-s11.2-g0.95 is quite different, because
its initial temperature profile differed significantly from those of
all other models.  At later times all models evolved quite
similarly. The early electron fraction distributions exhibit larger
difference than those of the other quantities, because the electron
capture rate strongly depends on temperature ($\propto T^6$), i.e., a
small difference in temperature can result in a large difference in
$Y_e$. However, once $\beta-$equilibrium is achieved, the $Y_e$
distributions of the models become quite similar (see dotted lines in
bottom panel).

\begin{figure}
\centering
\includegraphics[width=0.45\textwidth]{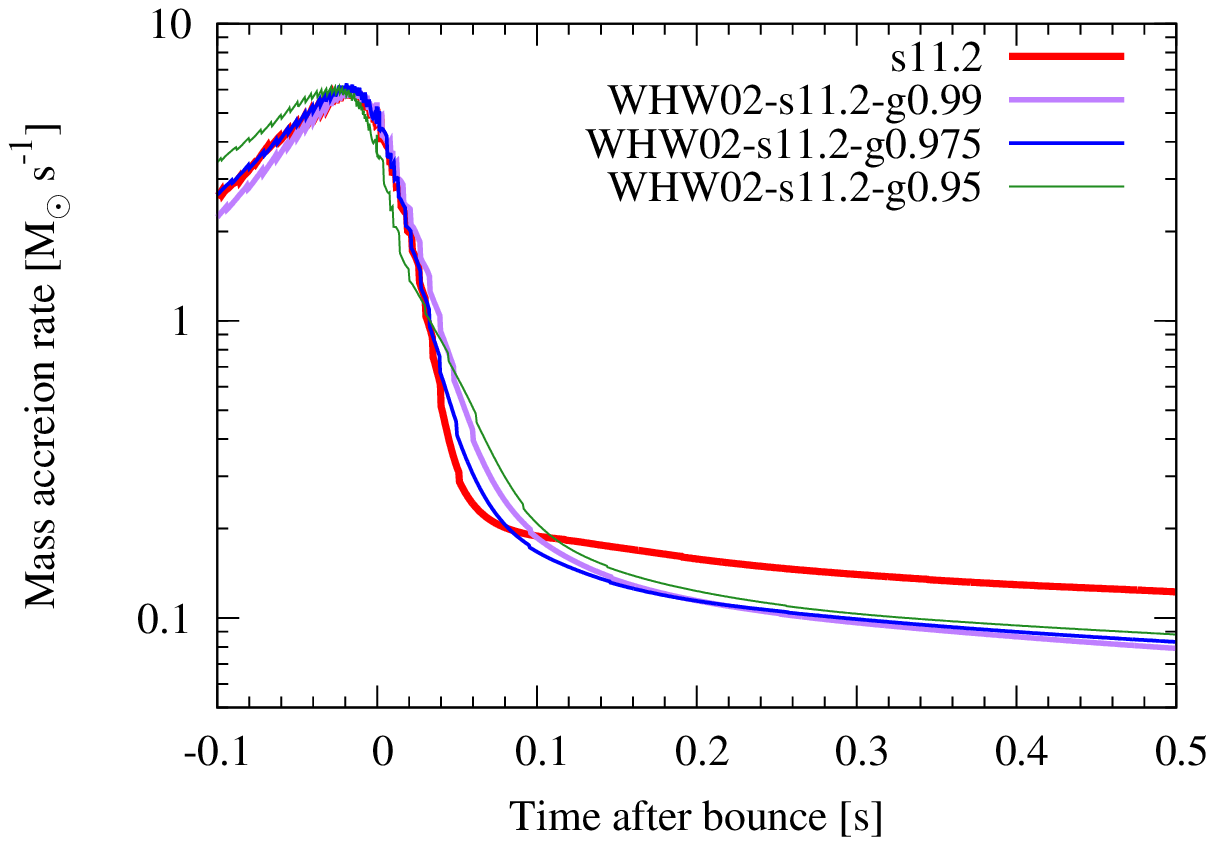}
\includegraphics[width=0.45\textwidth]{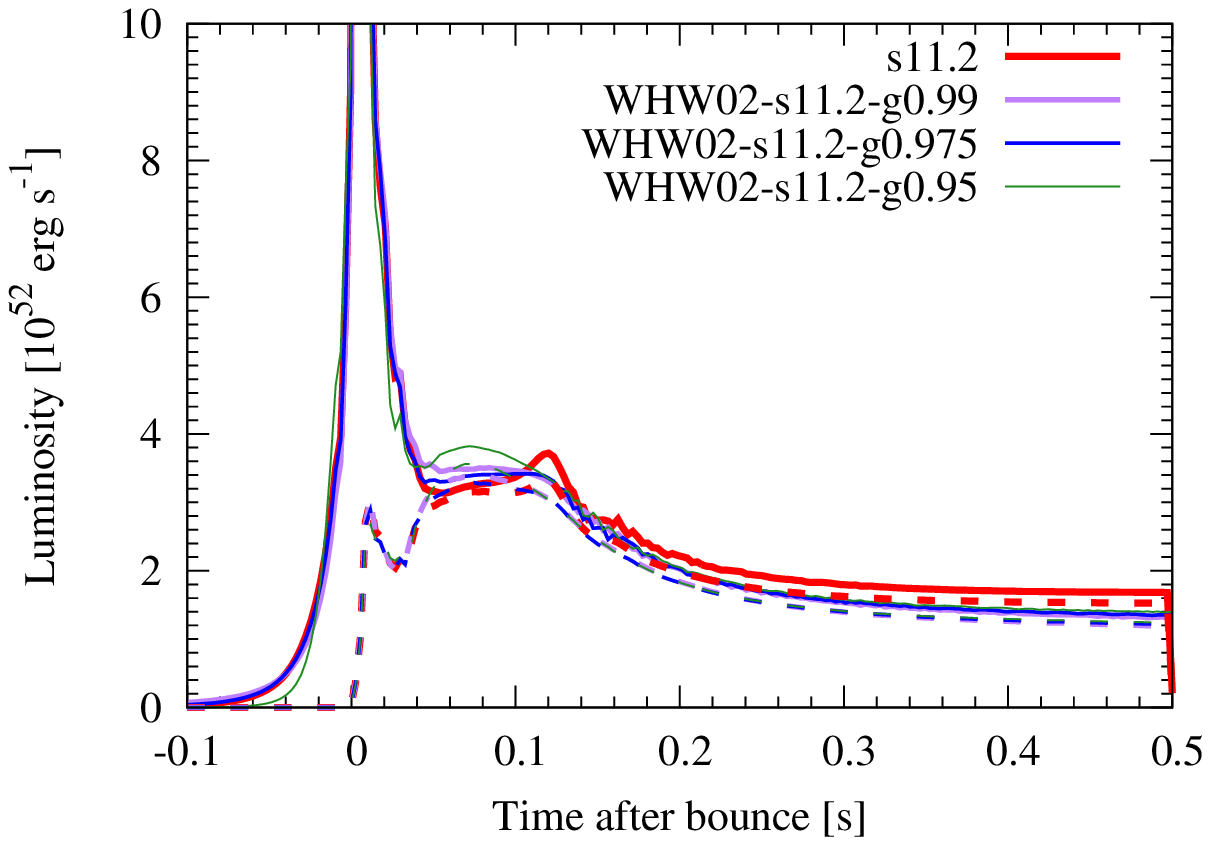}
\caption{Evolution of the mass accretion rate (top) and of the
  electron neutrino (solid) and electron antineutrino (dashed)
  luminosities (bottom) of the stellar evolutionary model s11.2 (red)
  and our three corresponding progenitor models (purple, blue, and
  green).}
\label{fig:mdot}
\end{figure}

In Fig.\,\ref{fig:mdot} we display the evolution of the mass accretion
rate measured at a radius of 300\,km (top panel) and of the electron
neutrino (solid lines) and electron antineutrino (dashed lines)
luminosities of the stellar evolutionary model s11.2 and of our three
corresponding progenitor models. Because of small differences in the
density structures of the models, both the mass accretion rates and
the neutrino luminosities differ slightly between the models. About
50\,ms post bounce, model s11.2 has the smallest mass accretion rate
because the density gradient at $M \approx 1.3 M_\odot$ is steepest
for this model. At later times ($\sim 100$\,ms post bounce) the mass
accretion rate is largest in this model, because its density is the
largest of all models in the relevant mass range $1.3 \lesssim
M/M_\odot \lesssim 1.5$ (see Fig.\,\ref{fig:density_s112}).

\begin{figure}
\centering
\includegraphics[width=0.45\textwidth]{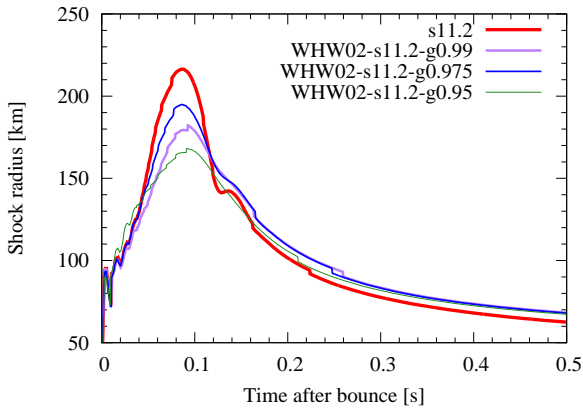}
\caption{Shock trajectories for the stellar evolutionary model s11.2
  (red) and our three corresponding progenitor models (purple, blue,
  and green). }
\label{fig:shock}
\end{figure}

Fig.\,\ref{fig:shock} illustrates the shock evolution after core
bounce. We define the shock position as the radius where the specific
entropy reaches a value of $6\,k_B$ baryon$^{-1}$. Model s11.2 has the
largest peak shock radius among all the investigated models, because
it possesses the steepest density gradient (see
Fig.\,\ref{fig:density_s112}). This fact leads to a rapid decrease of
the mass accretion rate with radius and hence of the ram pressure on
the shock. Since the shock radius is determined by the force balance
between the thermal post-shock pressure and the pre-shock ram
pressure, a lower ram pressure gives rise to a larger shock
radius. Our three other corresponding progenitor models also show
slightly different shock evolutions because their mass accretion rates
differ from each other and from that of model s11.2 (see
Fig.\,\ref{fig:mdot}).

\section{Parameter dependencies and explosion properties}
\label{sec:bc90}

\begin{table*}
\centering
\caption{Parameters characterizing the entropy, electron fraction, and
  density distributions of our initial models, which all had $M_1 =
  0.72 M_\odot$, $M_2 = 1.0 M_\odot$, $M_3 = 1.1 M_\odot$, $M_4 = 1.15
  M_\odot$, $M_5 = 1.17M_\odot$, $Y_{e4} =0.5$, and $g_\mathrm{eff} =
  0.975$. The parameter values which differ from those of model BC01
  are given in boldface. }
\begin{tabular}{lccccccc}
\hline
Model & $S_c$ & $S_1$ & $S_2$ & $S_5$       & $Y_{ec}$ & $Y_{e3}$ & $\rho_c$\\
      & \multicolumn{4}{c|}{[$k_B/$baryon]} & &   & [$10^{10}$\,g\,cm$^{-3}$] \\
\hline
BC01  & 0.5        & 0.63       & 1.6       & 4.0       & 0.415       & 0.46        & 2.0 \\ 
BC02  & {\bf 0.4}  & 0.63       & 1.6       & 4.0       & 0.415       & 0.46        & 2.0 \\ 
BC03  & {\bf 0.6}  & 0.63       & 1.6       & 4.0       & 0.415       & 0.46        & 2.0 \\ 
BC04  & 0.5        & {\bf 0.53} & 1.6       & 4.0       & 0.415       & 0.46        & 2.0 \\ 
BC05  & 0.5        & {\bf 0.73} & 1.6       & 4.0       & 0.415       & 0.46        & 2.0 \\ 
BC06  & 0.5        & 0.63       & {\bf 1.5} & 4.0       & 0.415       & 0.46        & 2.0 \\ 
BC07  & 0.5        & 0.63       & {\bf 1.7} & 4.0       & 0.415       & 0.46        & 2.0 \\ 
BC08  & 0.5        & 0.63       & 1.6       & {\bf 3.0} & 0.415       & 0.46        & 2.0 \\ 
BC09  & 0.5        & 0.63       & 1.6       & {\bf 6.0} & 0.415       & 0.46        & 2.0 \\ 
BC10  & 0.5        & 0.63       & 1.6       & 4.0       & {\bf 0.411} & 0.46        & 2.0 \\ 
BC11  & 0.5        & 0.63       & 1.6       & 4.0       & {\bf 0.425} & 0.46        & 2.0 \\ 
BC12  & 0.5        & 0.63       & 1.6       & 4.0       & 0.415       & {\bf 0.452} & 2.0 \\ 
BC13  & 0.5        & 0.63       & 1.6       & 4.0       & 0.415       & {\bf 0.47}  & 2.0 \\ 
BC14  & 0.5        & 0.63       & 1.6       & 4.0       & 0.415       & 0.46        & {\bf 1.0} \\ 
BC15  & 0.5        & 0.63       & 1.6       & 4.0       & 0.415       & 0.46        & {\bf 3.0} \\ 
\hline
BC16  & {\bf 0.4}  & {\bf 0.73} & 1.6       & 4.0       & 0.415       & 0.46       & 2.0 \\ 
BC17  & {\bf 0.4}  & 0.63       & {\bf 1.7} & 4.0       & 0.415       & 0.46       & 2.0 \\ 
BC18  & {\bf 0.4}  & 0.63       & 1.6       & {\bf 6.0} & 0.415       & 0.46       & 2.0 \\ 
BC19  & {\bf 0.4}  & 0.63       & 1.6       & 4.0       & {\bf 0.425} & 0.46       & 2.0 \\ 
BC20  & {\bf 0.4}  & 0.63       & 1.6       & 4.0       & 0.415       & {\bf 0.47} & 2.0 \\ 
BC21  & {\bf 0.4}  & 0.63       & 1.6       & 4.0       & 0.415       & 0.46       & {\bf 1.0} \\ 
BC22  & {\bf 0.4}  & 0.63       & 1.6       & 4.0       & 0.415       & 0.46       & {\bf 3.0} \\ 
\hline
\end{tabular}
\label{tab2}
\end{table*}

In the last section, we demonstrated the reliability of the new method
for constructing initial conditions for core-collapse supernova
simulations by comparing models constructed by this method with a
particular widely used presupernova model (WHW02-s11.2). The
hydrodynamic features of these models agree with each other quite
well. In Appendix \ref{sec:other}, we provide fitting parameters (see
Table\,A1) which closely approximate the density structures of other
presupernova models used in the literature (see
Fig.\,\ref{fig:other_prog}).

Next we consider a second set of initial conditions differing from
those reproducing progenitor models based on stellar evolutionary
calculations. In particular, we present our numerical results for
parameterized initial models based on model 109 of \cite{baro90},
which has a relatively small central entropy and a small core mass,
i.e. its structure differs significantly from that of initial models
obtained with stellar evolutionary calculations. Thus, this second set
of parametrized initial models allows us to study the dependence of
the outcome of core-collapse supernova simulations for quite different
initial conditions. The corresponding model parameters are given in
Table \ref{tab2}.

We first changed the value of one parameter from model to model (BC01
to BC15 in Table\,\ref{tab2}), and then we fixed the value of the
central entropy to $S_c=0.4$ and again changed one of the other
parameters from model to model(BC16 to BC22 in Table\,\ref{tab2}). As
we will show below, the reason for this approach was that model BC02
gives rise to a successful explosion, i.e., the parameter space around
this model is worth investigating. We note that we restricted the
parameters we chose in our study by the condition that the density at
$M_5$ is larger than $10^3$\,g\,cm$^{-3}$, which implies a lower limit
for the entropy or the electron fraction, because a low entropy or
electron fraction leads to a faster decrease of the density with
increasing mass coordinate.

\begin{table*}
\centering
\caption{Some properties characterizing our parametrized initial models}
\begin{tabular}{lcccccc}
\hline
Model & $R(M_5)^a$ & $\rho(M_5)^b$ & $T(M_5)^c$ & ${\xi_{M_5}}^d$ & ${\mu_{M_5}}^e$ & ${E_b}^f$ \\
 & [$10^8$ cm] & [$10^6$\,g\,cm$^{-3}$] & [$10^9$ K] &  &  & [B] \\
\hline
BC01 &  1.25 &   5.77 &   3.76 &   0.93 &   0.057 &   2.59  \\
BC02 &  1.50 &   1.98 &   2.73 &   0.78 &   0.028 &   2.50  \\
BC03 &  1.10 &   11.7 &   4.58 &   1.06 &   0.090 &   2.78  \\
BC04 &  1.81 &   0.53 &   1.78 &   0.65 &   0.011 &   2.47  \\
BC05 &  1.08 &   13.8 &   4.79 &   1.08 &   0.103 &   2.91  \\
BC06 &  1.44 &   2.13 &   2.80 &   0.81 &   0.028 &   2.50  \\
BC07 &  1.17 &   10.0 &   4.39 &   1.00 &   0.086 &   2.80  \\
BC08 &  1.22 &   7.29 &   3.44 &   0.96 &   0.069 &   2.52  \\
BC09 &  1.31 &   4.07 &   4.06 &   0.89 &   0.044 &   4.96  \\
BC10 &  1.72 &   0.81 &   2.05 &   0.68 &   0.015 &   2.47  \\
BC11 &  0.96 &   26.2 &   5.68 &   1.22 &   0.151 &   3.47  \\
BC12 &  1.98 &   0.27 &   1.41 &   0.59 &   0.007 &   2.47  \\
BC13 &  1.07 &   14.8 &   4.88 &   1.09 &   0.107 &   2.95  \\
BC14 &  1.56 &   3.14 &   3.15 &   0.75 &   0.048 &   2.61  \\
BC15 &  1.14 &   6.30 &   3.86 &   1.02 &   0.052 &   2.54  \\
\hline           
BC16 &  1.19 &   8.17 &   4.15 &   0.98 &   0.073 &   2.68  \\
BC17 &  1.29 &   5.68 &   3.75 &   0.90 &   0.060 &   2.62  \\
BC18 &  1.58 &   1.56 &   3.06 &   0.74 &   0.025 &   5.60  \\
BC19 &  1.01 &   19.2 &   5.24 &   1.16 &   0.125 &   3.14  \\
BC20 &  1.16 &   9.44 &   4.32 &   1.01 &   0.081 &   2.72  \\
BC21 &  1.90 &   0.91 &   2.13 &   0.61 &   0.021 &   2.49  \\
BC22 &  1.43 &   1.60 &   2.56 &   0.82 &   0.021 &   2.48  \\
     \hline
\end{tabular}
\begin{flushleft}
$^a$ Radius of $M=M_5$\\
$^b$ Density of $M=M_5$\\
$^c$ Temperature of $M=M_5$\\
$^d$ Compactness parameter, 
     $\xi_M\equiv(M/M_\odot)/[R(M)/1000~\mathrm{km}]$\\
$^e$ $\mu_M\equiv dM/dR=4\pi R(M)^2\rho(M)$ in units of 
     $M_\odot/1000$\,km\\
$^f$ Total binding energy\\
\end{flushleft}
\label{tab3}
\end{table*}

In Table\,\ref{tab3} we give the values of some quantities
characterizing the density structures of our second set of
parametrized models.  Columns 2 to 4 give the radius (in units of
$10^8$\,cm), the density (in units of $10^6$\,g\,cm$^{-3}$), and the
temperature (in units of $10^9$\,K) at the mass coordinate $M=M_5$,
respectively. In the fifth column we list the compactness parameter
$\xi_M$ \citep{ocon11}, which is defined as
\begin{equation}
 \xi_M = \frac{M/M_\odot}{R(M)/1000\,\mathrm{km}},
\end{equation}
where $R(M)$ is the radius of the sphere containing a mass $M$.  Note
that we use here the compactness parameter $\xi_{M_5}$, whereas
\cite{ocon11} considered $\xi_{M=2.5M_\odot}$ instead. According to
\cite{ocon11} smaller values of $\xi_M$ are better for explosions.
Column 6 gives the parameter $\mu_M$, defined by \citet{ertl16} as
\begin{equation}
 \mu_M = \left.\frac{dM}{dr}\right|_{r=R(M)}=4\pi \rho R^2(M).
\label{eq:mu_M}
\end{equation}
Whereas \citet{ertl16} obtained the value of $dM/dr$ by computing the
numerical derivative of $dM/dr$ at the mass shell where
$S=4k_B$\,baryon$^{-1}$ with a mass interval of $0.3M_\odot$, we used
for simplicity the second equality in Eq.\,(\ref{eq:mu_M}) to compute
$dM/dr$ analytically.  \citet{ertl16} showed that for a given value of
$M_{S=4}$ (the mass coordinate where $S=4k_B$\,baryon$^{-1}$), a
smaller value of $\mu_M$ is better for an explosion.  Finally, the
last column gives the total binding energy of the initial model, which
includes the contribution of the internal energy.

\begin{figure}
\includegraphics[width=0.45\textwidth]{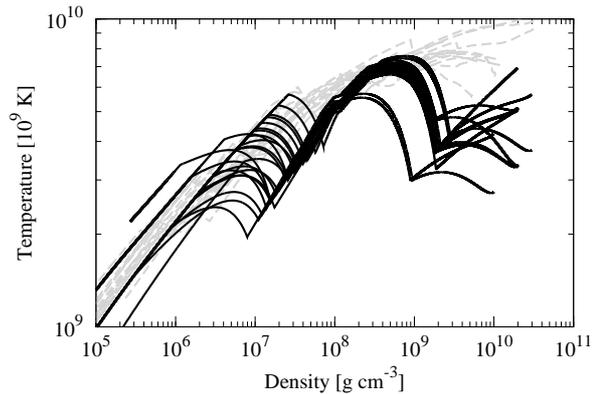}
\caption{Model structure in a temperature-density plane. Black and
  grey lines give the profiles of our parameterized models and of
  models obtained from stellar evolutionary calculations (see Appendix
  \ref{sec:other}), respectively.}
\label{fig:d-t_BC90}
\end{figure}

Fig.\,\ref{fig:d-t_BC90} shows the structure of our second set of
parameterized models (solid lines) in a density-temperature plane. The
additional grey lines give the structures of the models listed in
Appendix \ref{sec:other}, which are obtained by stellar evolution
calculations. Obviously, our parametrized models show a similar trend
as the evolutionary ones, except for their non-monotonic behavior at
densities $\rho\sim 10^7$ g cm$^{-3}$ and at densities of a few times
$10^9$ g cm$^{-3}$, i.e. near the center.  In other words, these
models allow us to investigate thermodynamic regimes beyond those
encountered in canonical models.

The Chandrasekhar mass is often used as a rough estimate of the iron
core mass. Since the former mass depends on the electron fraction as
\begin{align}
M_{ch}& \approx 5.87 Y_e^2 M_\odot\\
      & =1.01 \left( \frac{Y_e}{0.415} \right)^2M_\odot,
\end{align}
our small iron core ($M_4=1.15M_\odot$) can be unstable.

\begin{table*}
\centering
\caption{Overview of our hydrodynamic simulations}
\begin{tabular}{lcccccccccc}
\hline
Model & ${t_\mathrm{bounce}}^a$ & ${t_{400}}^b$  & ${t_\mathrm{fin}}^c$ & ${R_\mathrm{shock,max}}^d$ & ${M_\mathrm{NS,final}}^e$ & ${E_\mathrm{diag,1000}}^f$ & ${E_\mathrm{diag,fin}}^g$ & ${\dot{E}_\mathrm{diag,fin}}$$^h$ & ${Y_{c,\mathrm{bounce}}}^i$ & ${E_\mathrm{kin,init}}^j$\\
& [ms] & [ms] & [ms] & [$10^8$ cm] & [$M_\odot$] & [B] & [B] & [B s$^{-1}$] & & [B] \\
\hline
BC01 & 252 &  --- &  944 & 0.244 & 1.22 &   --- &   --- &   --- & 0.328 & 7.81  \\
BC02 & 254 & 21.9 &  452 & 2.908 & 1.09 & 0.145 & 0.294 &   3.8 & 0.336 & 8.44  \\
BC03 & 245 &  --- &  981 & 0.203 & 1.27 &   --- &   --- &   --- & 0.319 & 7.09  \\
BC04 & 248 &  --- &  841 & 0.274 & 1.19 &   --- &   --- &   --- & 0.327 & 7.80  \\
BC05 & 255 &  --- &  928 & 0.209 & 1.28 &   --- &   --- &   --- & 0.328 & 7.71  \\
BC06 & 242 &  --- & 1000 & 0.350 & 1.19 &   --- &   --- &   --- & 0.328 & 7.84  \\
BC07 & 261 &  --- &  953 & 0.214 & 1.26 &   --- &   --- &   --- & 0.327 & 7.83  \\
BC08 & 242 &  --- &  833 & 0.232 & 1.20 &   --- &   --- &   --- & 0.328 & 7.92  \\
BC09 & 252 &  --- & 1000 & 0.312 & 1.26 &   --- &   --- &   --- & 0.327 & 7.76  \\
BC10 & 249 &  --- &  855 & 0.316 & 1.18 &   --- &   --- &   --- & 0.327 & 7.65  \\
BC11 & 249 &  --- & 1000 & 0.194 & 1.37 &   --- &   --- &   --- & 0.327 & 7.82  \\
BC12 & 239 &  --- &  709 & 0.279 & 1.17 &   --- &   --- &   --- & 0.327 & 7.66  \\
BC13 & 262 &  --- &  940 & 0.205 & 1.29 &   --- &   --- &   --- & 0.328 & 8.00  \\
BC14 & 401 &  --- &  997 & 0.248 & 1.21 &   --- &   --- &   --- & 0.327 & 7.68  \\
BC15 & 189 &  --- &  733 & 0.259 & 1.21 &   --- &   --- &   --- & 0.327 & 7.84  \\
\hline
BC16 & 259 &  --- & 1000 & 0.283 & 1.25 &   --- &   --- &   --- & 0.335 & 8.44  \\
BC17 & 263 & 23.8 &  496 & 2.487 & 1.10 & 0.160 & 0.331 &   3.4 & 0.336 & 8.39  \\
BC18 & 259 & 21.2 &  451 & 3.000 & 1.08 & 0.132 & 0.386 &   4.7 & 0.336 & 8.50  \\
BC19 & 254 &  --- & 1000 & 0.873 & 1.32 &   --- &   --- &   --- & 0.340 & 9.50  \\
BC20 & 267 &  --- & 1000 & 0.542 & 1.26 &   --- &   --- &   --- & 0.336 & 8.51  \\
BC21 & 397 & 22.5 &  590 & 3.120 & 1.09 & 0.090 & 0.269 &   4.1 & 0.336 & 8.45  \\
BC22 & 192 & 22.2 &  392 & 3.060 & 1.08 & 0.141 & 0.234 &   2.9 & 0.336 & 8.48  \\
\hline
\end{tabular}
\begin{flushleft}
$^a$ Time until bounce since the beginning of the simulation\\
$^b$ Time past bounce when the shock reaches a radius of 400\,km 
     in the exploding models \\
$^c$ Final time of the simulation \\
$^d$ Maximum shock radius \\
$^e$ Final mass of the PNS, which is defined by $\rho > 10^{11}$\,g\,cm$^{-3}$ \\
$^f$ Diagnostic explosion energy when the shock reaches a radius of 1000\,km \\
$^g$ Diagnostic explosion energy at $t_\mathrm{fin}$, when it is still increasing \\
$^h$ Growth rate of the diagnostic explosion energy estimated 30\,ms 
      before $t_\mathrm{fin}$ \\
$^i$ Value of $Y_e$ in the center at $t_\mathrm{bounce}$ \\
$^j$ Initial kinetic energy, which is estimated by the maximum value
  of the kinetic energy inside the mass of the largest infall velocity\\
\end{flushleft}
\label{tab4}
\end{table*}

\begin{figure}
\includegraphics[width=0.45\textwidth]{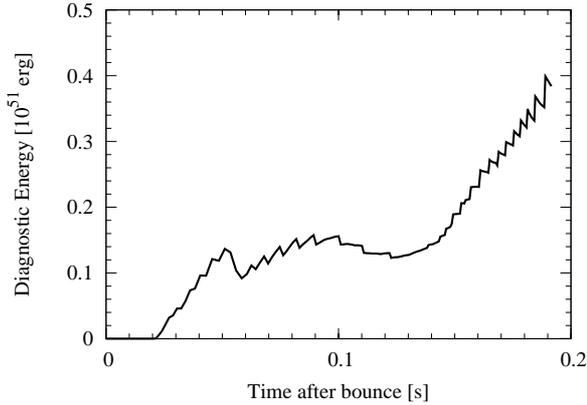}
\caption{Evolution of the diagnostic explosion energy for model BC18.}
\label{fig:exp_ene}
\end{figure}

In Table\,\ref{tab4} we provide an overview of the hydrodynamic
simulations with our second set of models. The table columns give the
time until bounce, the postbounce time when the shock reaches a radius
of 400 km, the final time of the simulation, the maximum shock radius,
the final baryonic mass of the PNS, and the diagnostic explosion
energy at the times when the shock reaches a radius of 1000km and at
${t_\mathrm{fin}}$, respectively. The growth rate of the diagnostic
energy at ${t_\mathrm{fin}}$ is given in the next column.  The
remaining columns give the value of $Y_e$ in the center at
$t_\mathrm{bounce}$, and the initial kinetic energy.  The PNS mass is
defined as the mass with $\rho>10^{11}$ g cm$^{-3}$, and the
diagnostic explosion energy as the integral of the local energy,
i.e. the sum of the specific internal, kinetic and gravitational
energies, of all zones where this quantity and the radial velocity are
positive. Here we used the general relativistic expression for the
local energy of \cite{muel12b}, which is given as
\begin{equation}
 e_\mathrm{local} = \alpha \left[ \left(\rho c^2+\epsilon c^2+P\right)
   W^2-P \right] - \rho W c^2,
\end{equation}
where $\alpha$ is the lapse function, $c$ the speed of light,
$\epsilon$ the specific internal energy, and $W$ the Lorentz
factor. This expression reduces to the well-known Newtonian expression
($e_\mathrm{local} = \rho\phi + \rho v^2/2 + \rho\epsilon$ with $\phi$
and $v$ being the gravitational potential and the velocity,
respectively) when one omits higher-order terms like $(v/c)^2$.

\begin{figure}
\includegraphics[width=0.45\textwidth]{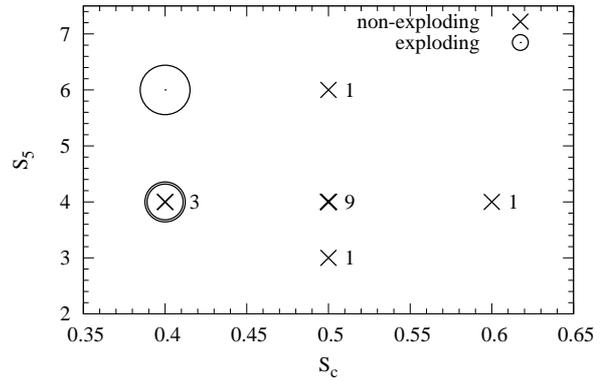}
\caption{Score sheet providing an overview (in the $S_c$-$S_5$ plane)
  of the models exploding or failing.  Exploding models are marked by
  open circles and non-exploding oners are represented by crosses. The
  size of the circle is proportional to the growth rate of the
  diagnostic explosion energy (see Table \ref{tab4}). The numbers
  attached to the symbols give the number of non-exploding models with
  $\rho_c=2\times 10^9$ g cm$^{-3}$.}
\label{fig:sc-s5}
\end{figure}

For model BC18, which produces the most energetic explosion of our
second set of models, $S_c=0.4$ and $S_5=6\,k_B$\,baryon$^{-1}$.  The
diagnostic explosion energy of this model already reaches 0.39\,B (=
$3.9\times 10^{50}$\,erg) at the end of the simulation, and it is
still increasing (see Fig.\,\ref{fig:exp_ene}) at a rate of
$\approx$4.7\,B\,s$^{-1}$, i.e., it will reach a value of 1\,B about
320\,ms after core bounce.
Figure \ref{fig:sc-s5} presents a score sheet, which provides an
overview (in the $S_c$-$S_5$ plane) of the models exploding or
failing. The exploding models are marked by open circles, while the
non-exploding models are represented by crosses. The size of the
circles is proportional to the growth rate of the diagnostic explosion
energy towards the end of the simulation (see Table
\ref{tab4}). Obviously, smaller values of $S_c$ lead to a larger
diagnostic explosion energy, and larger values of $S_5$ give rise to a
faster growth rate of the diagnostic explosion energy in case of the
exploding models.

Concerning the explosion energy one should note that the envelope
located above the Si/O layer has a large binding energy of
$O(10^{49})$ to $O(10^{51})$, the actual value depending on the ZAMS
mass of the progenitor \citep[e.g.][]{pejc15}. Therefore, the values
given in Table\,\ref{tab4} are not the observable explosion
energies. To determine the latter energies, one needs to perform
long-term simulations including the stellar envelopes, which will be
left for future work.

For our second set of models, the electron fraction at bounce is
larger than in the simulations with our first set of models based on
the stellar evolutionary model s11.2 (see previous section and
Fig.\,\ref{fig:s11.2}), in which $Y_{c,\mathrm{bounce}} \approx 0.3$.
Because of their smaller initial central entropy, the latter models
have a lower temperature, which implies a smaller electron capture
rate during collapse. The resulting larger electron fractions explain
the larger kinetic energies at the bounce \citep[see, e.g.][]{muel98},
which are given by the kinetic energy of the inner core at the ``last
good homology'' \citep{brow82}. Of the models listed in the upper part
of Table\,\ref{tab4}, model BC02 has the largest initial kinetic
energy.  Among these models, model BC02 is also the only exploding
model.  Although a higher value of $Y_{ec}$ also leads to a larger
value of $Y_{c,\mathrm{bounce}}$ and a larger initial kinetic energy
(see model BC19 in the lower part of Table\,\ref{tab4}), model BC19
does not explode because of its larger gravitational binding energy
(see Table\,\ref{tab3}). However, we note that in comparison to the
other non-exploding models (BC16-BC18, BC20-BC22), the shock
propagates out to an exceptionally large maximum shock radius of
873\,km in model BC19, i.e, it is a marginal model marking the
boundary between exploding and non-exploding models.

In all exploding models the explosion sets in early ($\sim$ 20\,ms
after core bounce), which seems to suggest a prompt
explosion. However, these explosions are still aided by neutrino
heating, i.e., they differ from prompt explosion models, in which
initial kinetic energy is large enough to eject the envelope. To
validate this statement, we performed a simulation without neutrino
heating by setting the distribution function of streaming particles,
which is essential for neutrino heating in IDSA \citep[see][]{lieb09},
to zero. Then, the exploding model does no longer explode, i.e., it
was no prompt explosion.

From these result, we conclude that the iron core structure is crucial
for obtaining an explosion. Especially, a low entropy at the center
helps to make an explosion. To reach a more general conclusion, we
need a large number of simulations covering a wider range of parameter
space, which will be reported in a forthcoming publication.

\section{Summary and discussion}
\label{sec:summary}

In this paper, we investigated a method to construct parametrized
initial progenitor models for core-collapse supernova simulations. So
far, initial conditions of these simulations have been taken from the
final phase of stellar evolutionary calculations, which depend on
several uncertainties, like the treatments of semi-convection,
overshooting, and composition mixing. In particular, many
phenomenological treatments are employed to approximate
multi-dimensional effects in evolution simulations done in spherical
symmetry.
\footnote{Recently, the question whether aspherical fluid motion can
  help the explosion has been the focus of several studies (see, e.g.,
  \cite{couc13c,muel15a,couc15b}).}
In this paper, we proposed a alternative methodology to construct
initial conditions. This is not a completely new idea and we reused a
method by \cite{baro90}. However, different from these authors, we
used the latest input microphysics including neutrino transfer, a
microscopic nuclear equation of state and general relativistic
hydrodynamics. In their paper, they presented functions of entropy and
electron fraction expressed by mass coordinate. With these functions
and solving hydrostatic equation, we can construct initial density
structures.

First of all, we constructed structures with parameters fitting the
commonly used model s11.2 from \cite{woos02} and showed the similarity
between our models and the model s11.2. We then performed general
relativistic neutrino-radiation hydrodynamics simulations in spherical
symmetry with the public code Agile-IDSA \citep{lieb09}
\footnote{The code is available from
  https://physik.unibas.ch/\~{}liebend/download/}
and showed the reliability of our method. Next, we constructed models
based on parameters given in \cite{baro90} and studied parameter
dependencies. Interestingly, we found several exploding models with
small central entropy even in spherically symmetric simulations. More
surprisingly, models with a large entropy in the Si/O layer give
rather large explosion energies, $\sim 4\times 10^{50}$ erg at the
final time of our simulations, and the energy still being increasing.

Nevertheless of the large explosion energy ($\gtrsim 4\times 10^{50}$
erg), we find that the PNS masses are rather small ($\sim 1.1 M_\odot$
in baryonic mass) for exploding models, so that these explosions are
not fully compatible with observations. This discrepancies will be
reduced when we use multi-dimensional simulations, since
multi-dimensional effects amplify neutrino heating and explodability
significantly. These simulations would produce an explosion for models
that do not explode in spherically-symmetric simulations for a larger
value of the central entropy, and would lead to continuous mass
accretion onto a PNS. With multi-D simulations, we may find parameter
sets leading to a large explosion energy, $\sim 10^{51}$ erg, and a
typical NS gravitational mass, $\sim$ 1.3--2.0 $M_\odot$, which is
typically $\sim 10$\% smaller than the baryonic mass. A broader
parameter survey is necessary to explore these more promising
combinations.

In this study we considered initial modesl with different entropy
stratificationss, but did not pay much attention to the temperature
profiles.  However, the temperature distribution is crucially
important for nuclear synthesis and the energy generation rate in
burning layers. Therefore, our current model might not be fully
consistent with stellar evolution and an improvement will be presented
in the forthcoming papers.

This work is the very first step toward investigating initial
conditions other than those resulting from stellar evolutionary
calculations. The virtue of the method used in our study is that we
can choose initial conditions beyond those predicted by current
stellar evolutionary calculations.  Hence, we may be able to find
robust conditions for energetic explosions, i.e. explosions in which
the energy is larger than the canonical value $10^{51}$ erg. This is
one of the important goals for the core-collapse supernova simulation
community.

\section*{Acknowledgements}

We thank M.\,Liebend\"orfer for providing Agile-IDSA and his routines
for producing EOS table to us, and A.\,Heger, M.\,Limongi, K.\,Nomoto,
S.\,Woosley, and T.\,Yoshida for providing data of pre-collapse
cores. YS thanks the Max Planck Institute for Astrophysics for its
hospitality. YS was supported by JSPS postdoctoral fellowships for
research abroad, MEXT SPIRE, and JICFuS. EM is partially supported by
the Cluster of Excellence EXC 153 ``Origin and Structure of the
Universe''\footnote{http://www.universe-cluster.de}.

\appendix

\section{Fitting other progenitor models}
\label{sec:other}

\begin{figure*}
\includegraphics[width=0.32\textwidth]{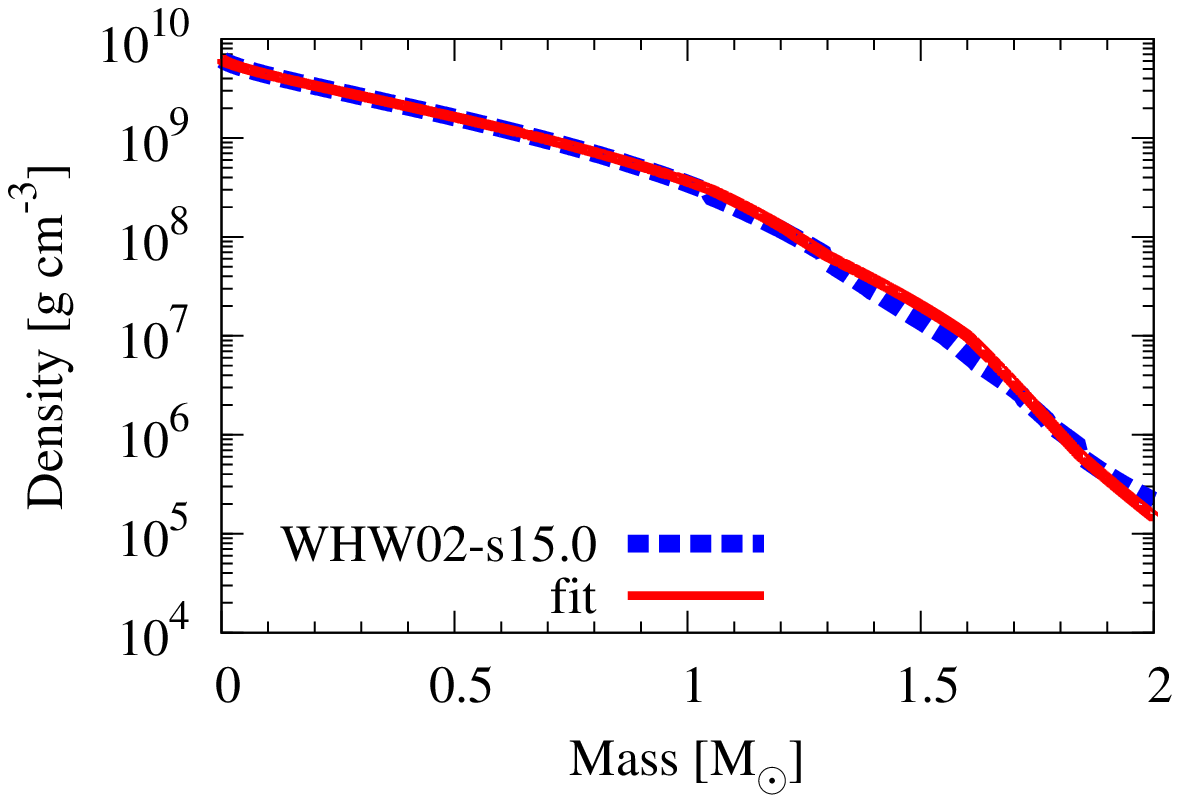}
\includegraphics[width=0.32\textwidth]{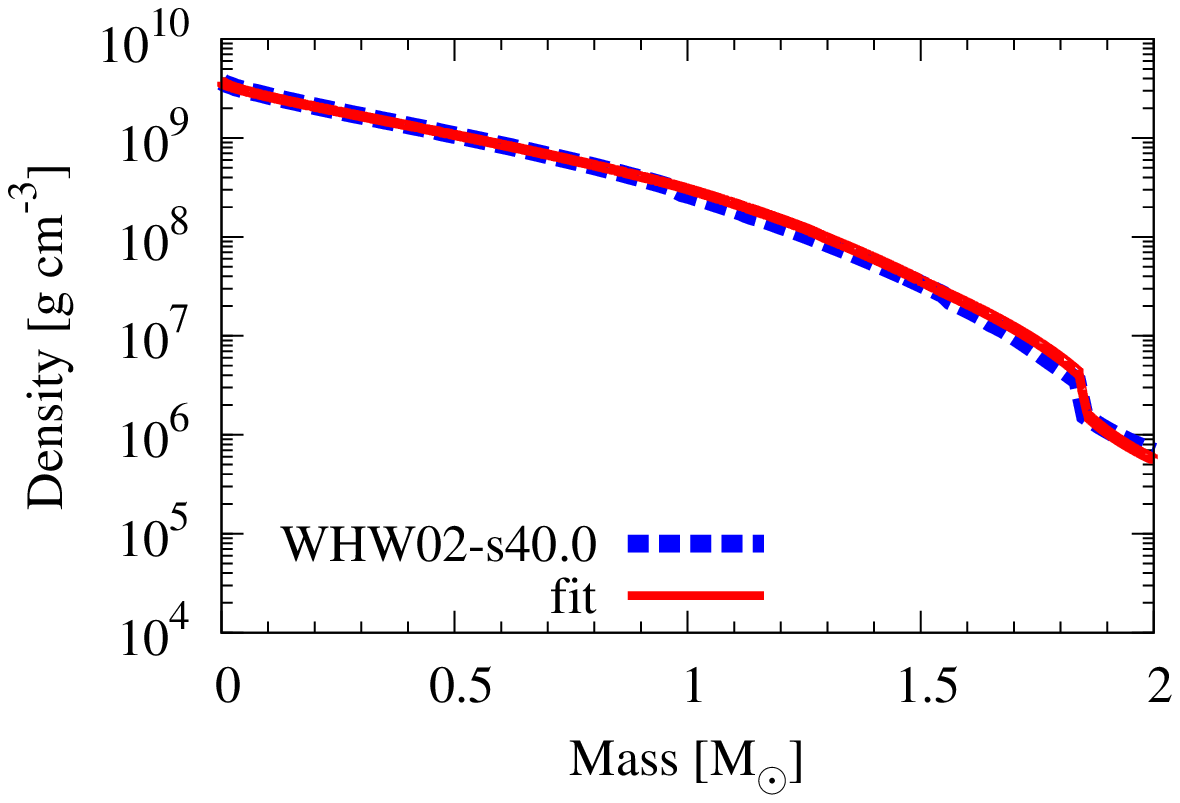}
\includegraphics[width=0.32\textwidth]{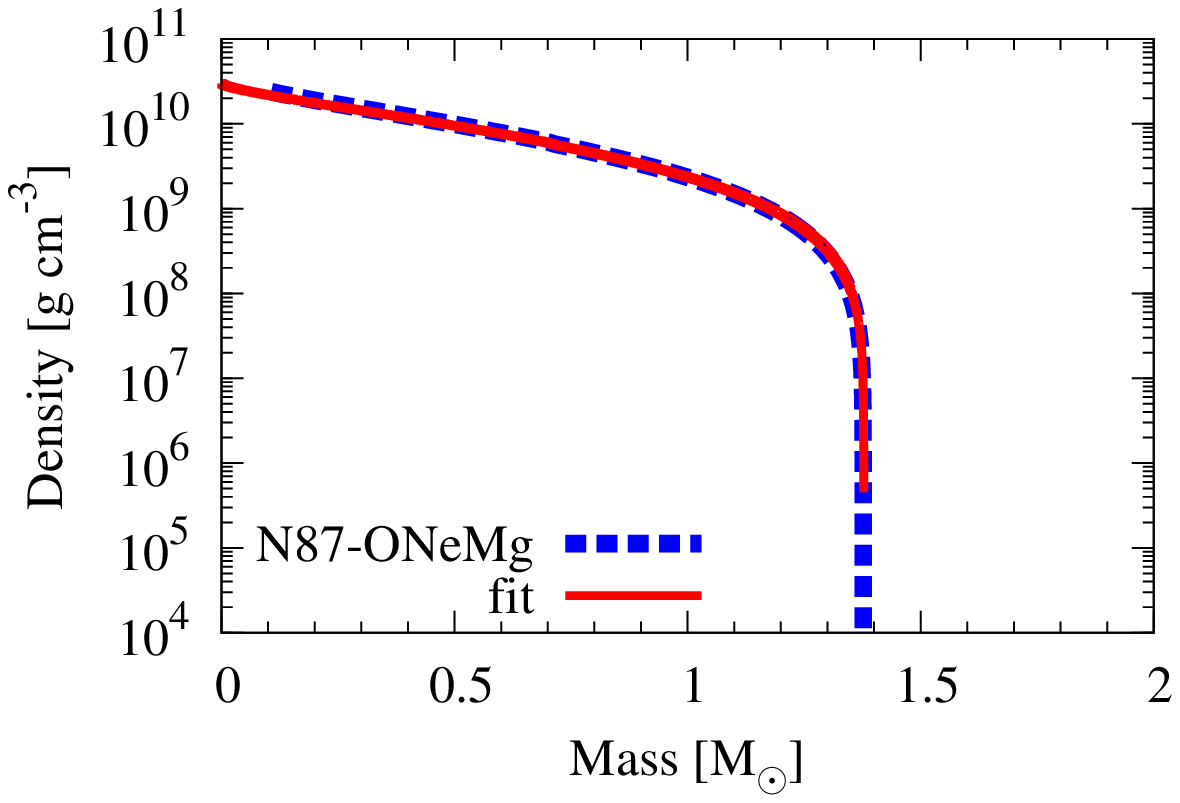}
\includegraphics[width=0.32\textwidth]{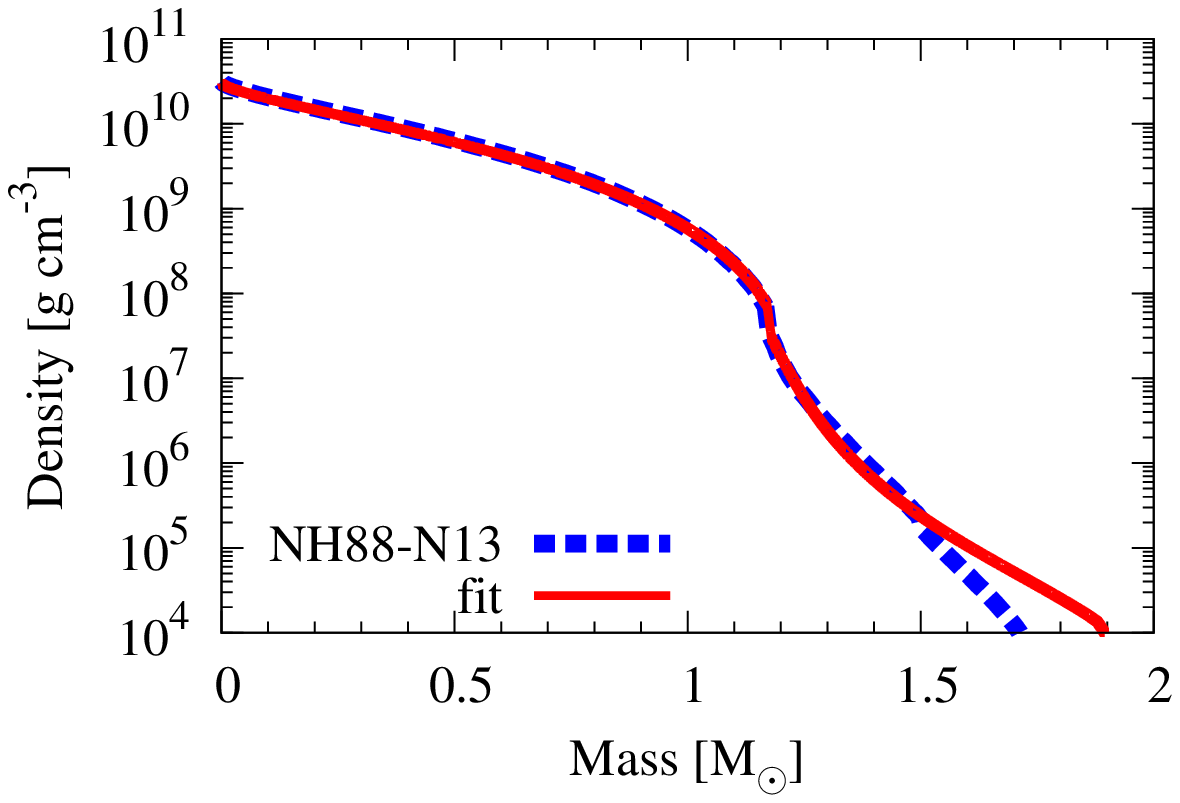}
\includegraphics[width=0.32\textwidth]{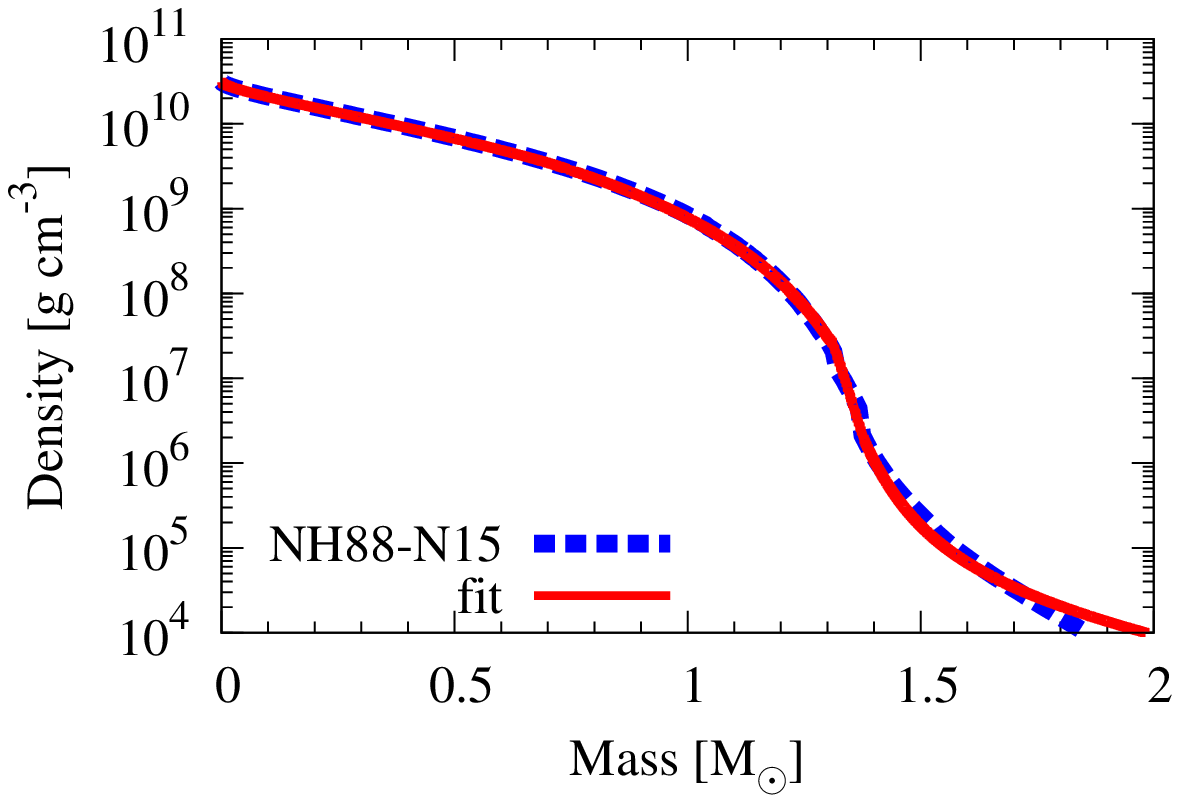}
\includegraphics[width=0.32\textwidth]{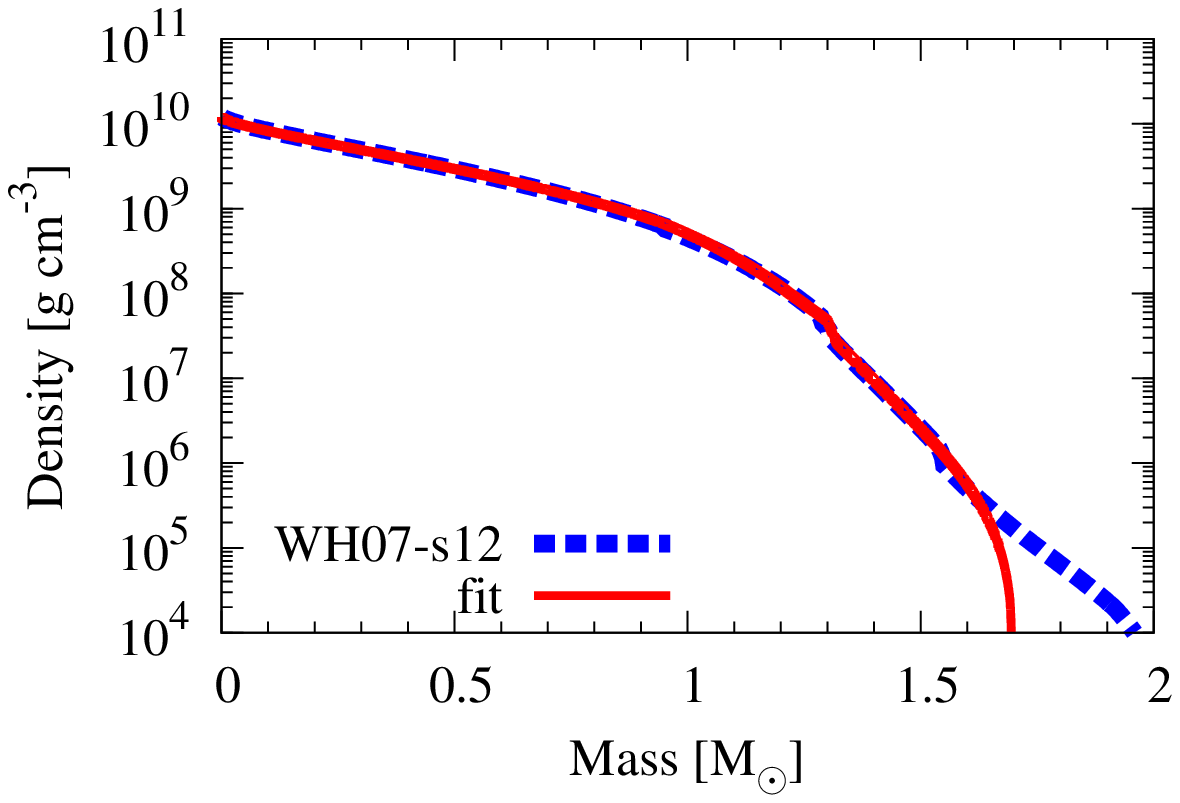}
\includegraphics[width=0.32\textwidth]{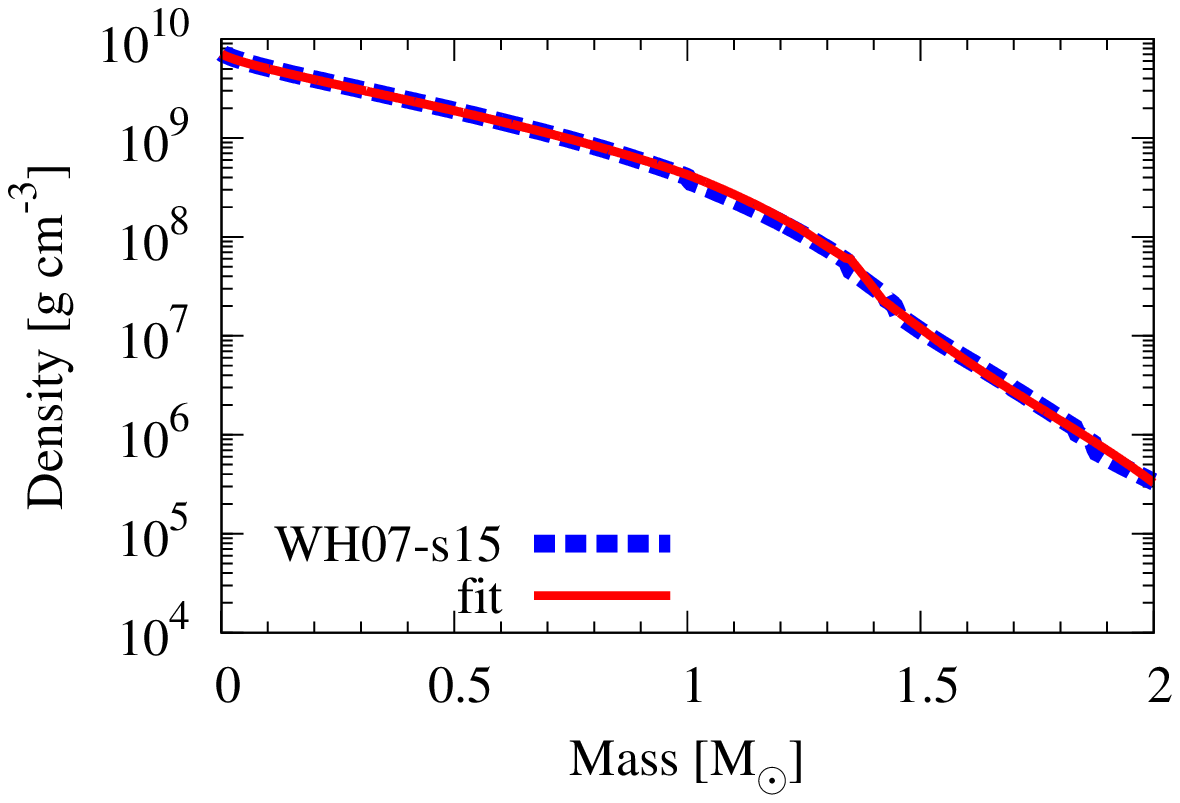}
\includegraphics[width=0.32\textwidth]{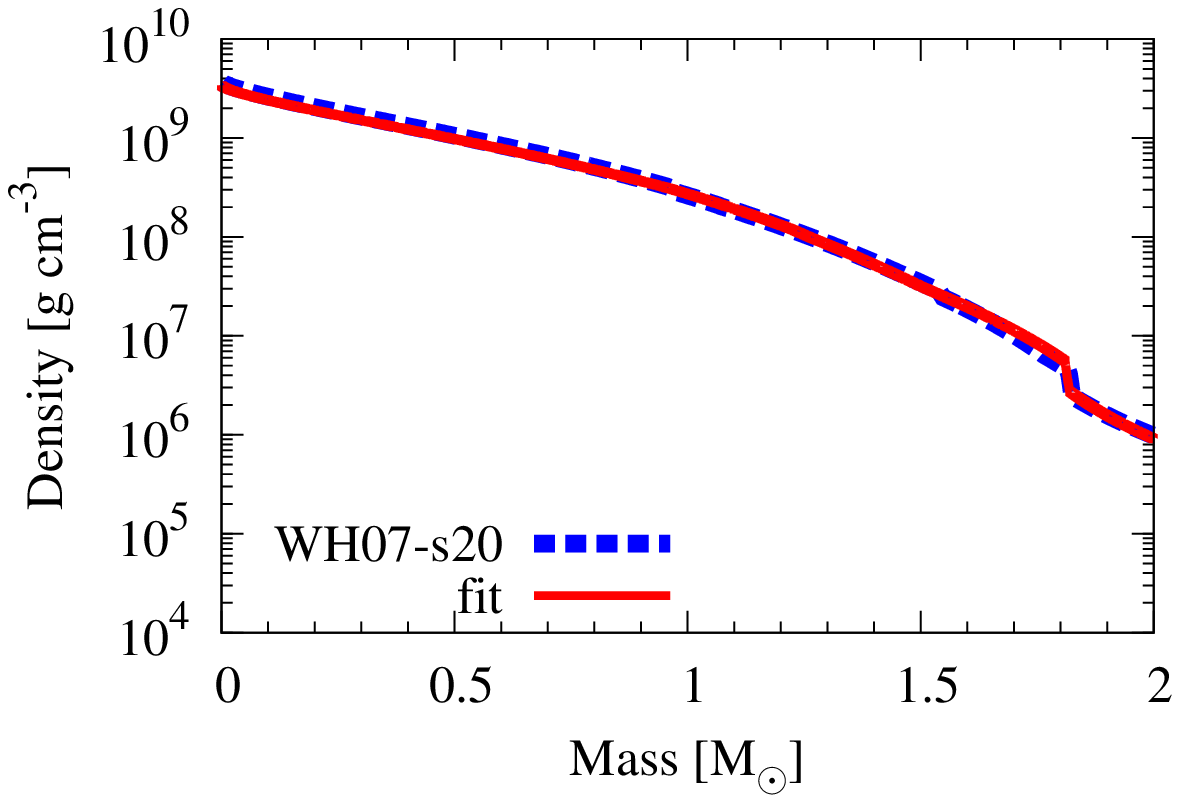}
\includegraphics[width=0.32\textwidth]{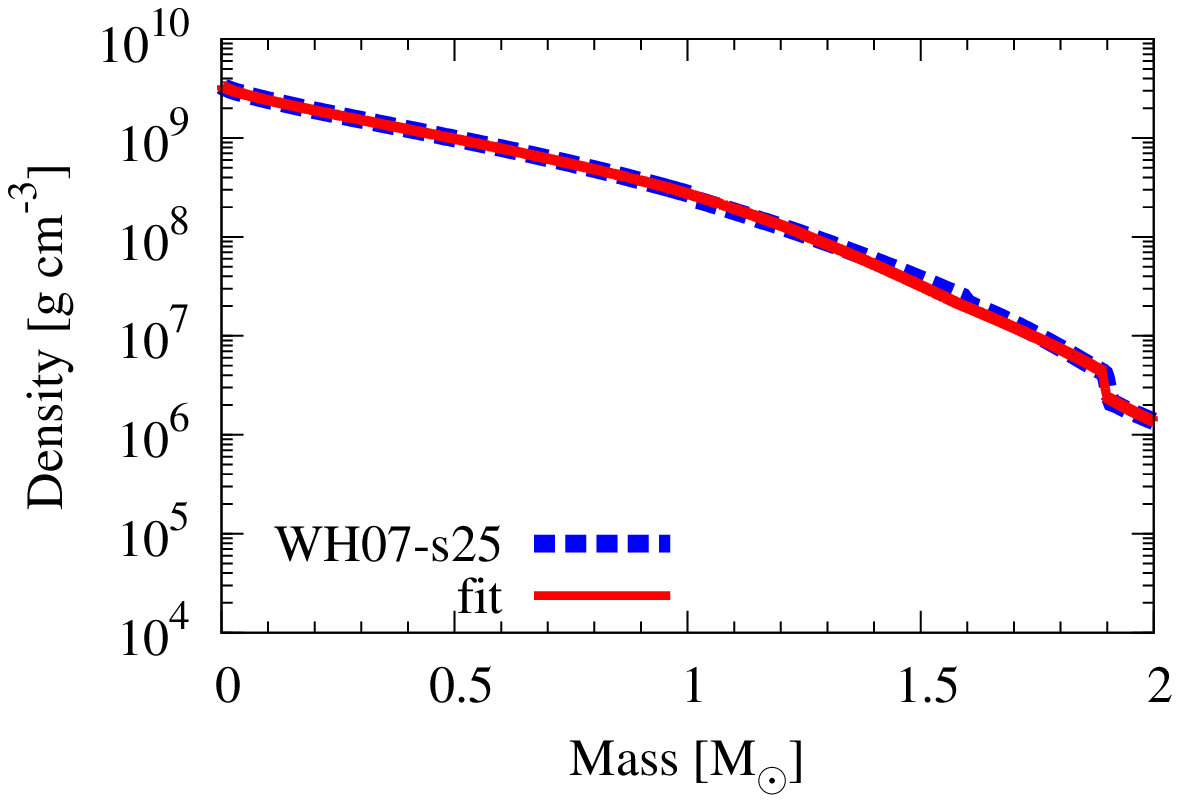}
\includegraphics[width=0.32\textwidth]{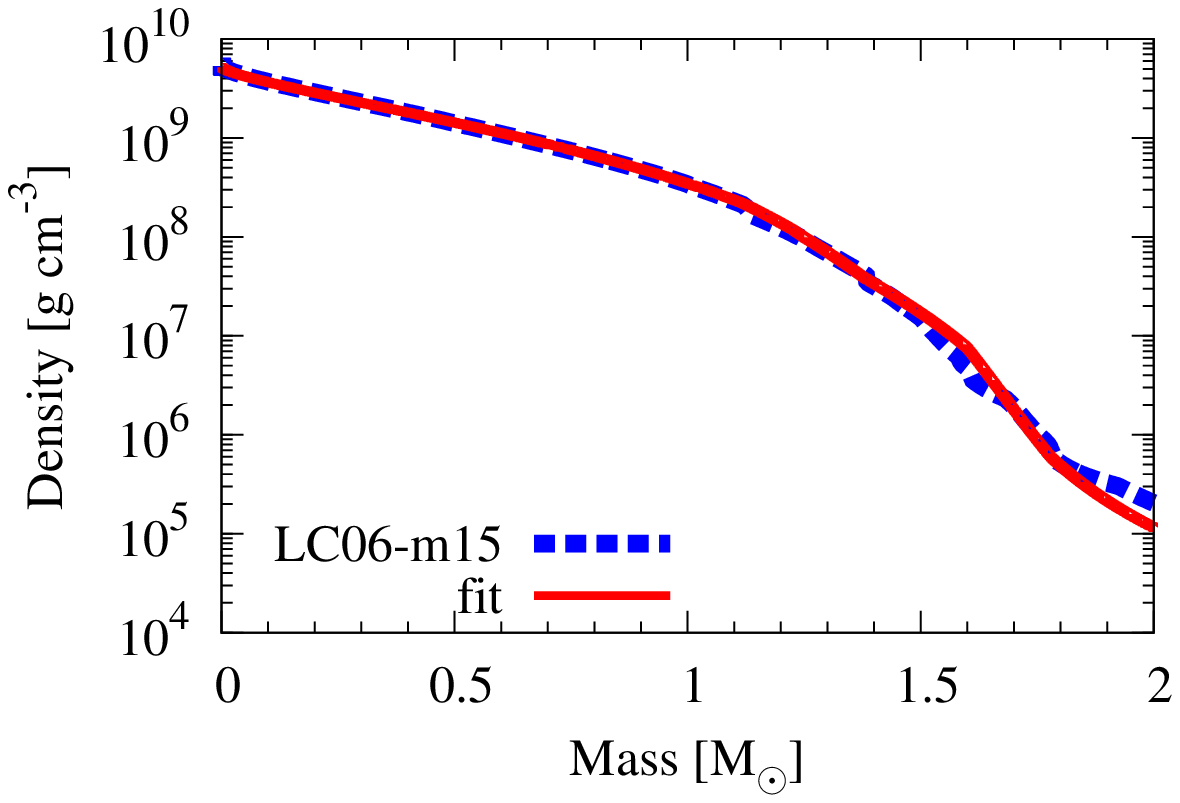}
\includegraphics[width=0.32\textwidth]{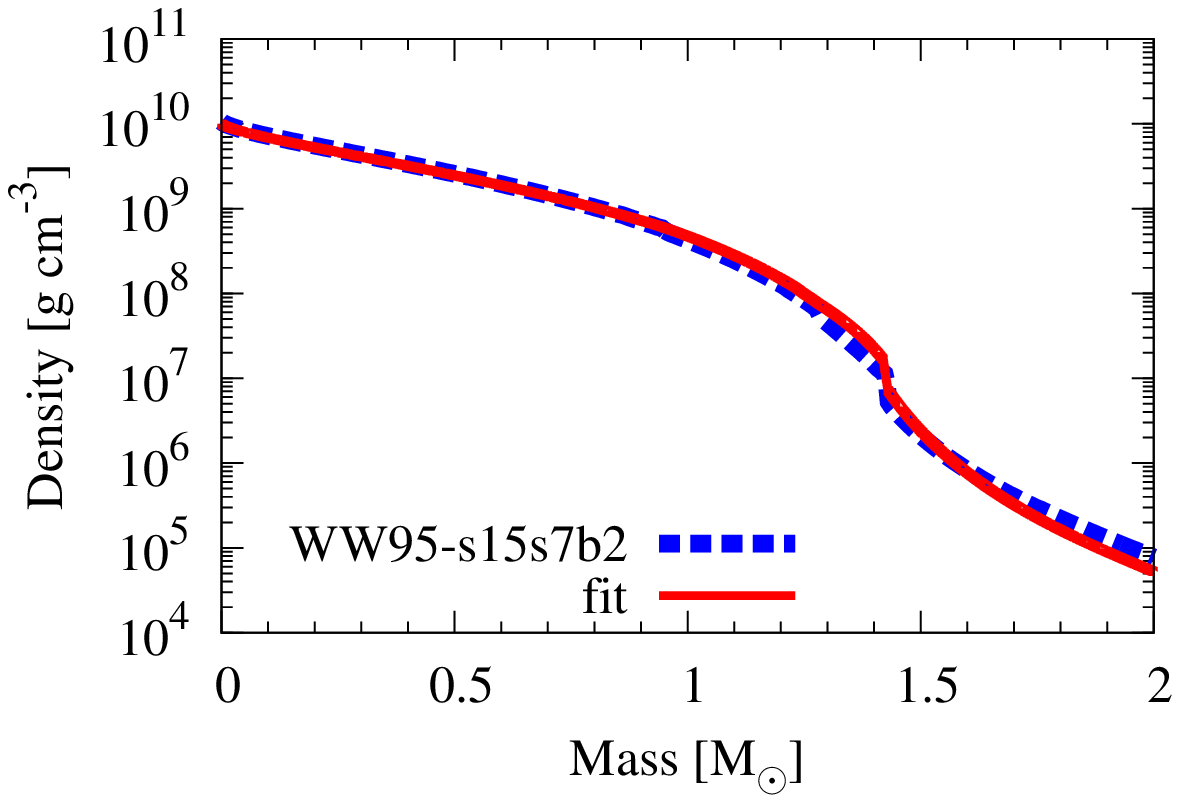}
\includegraphics[width=0.32\textwidth]{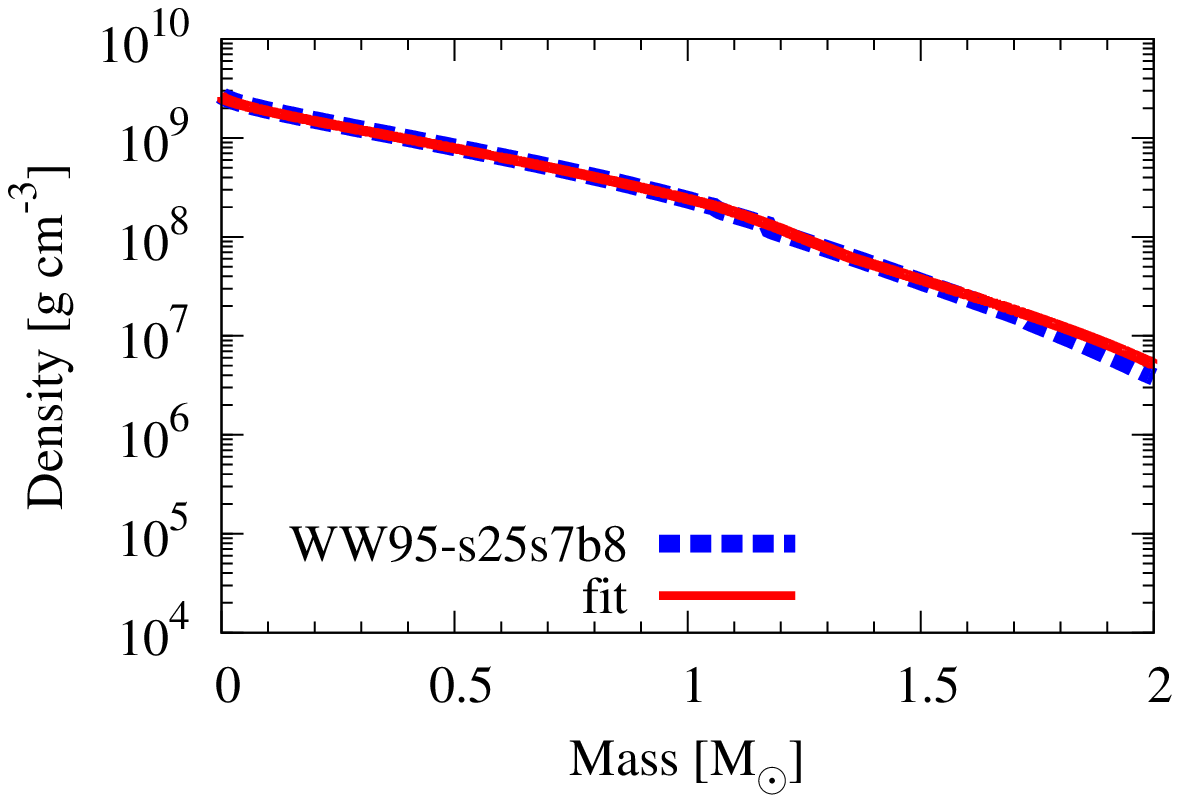}
\includegraphics[width=0.32\textwidth]{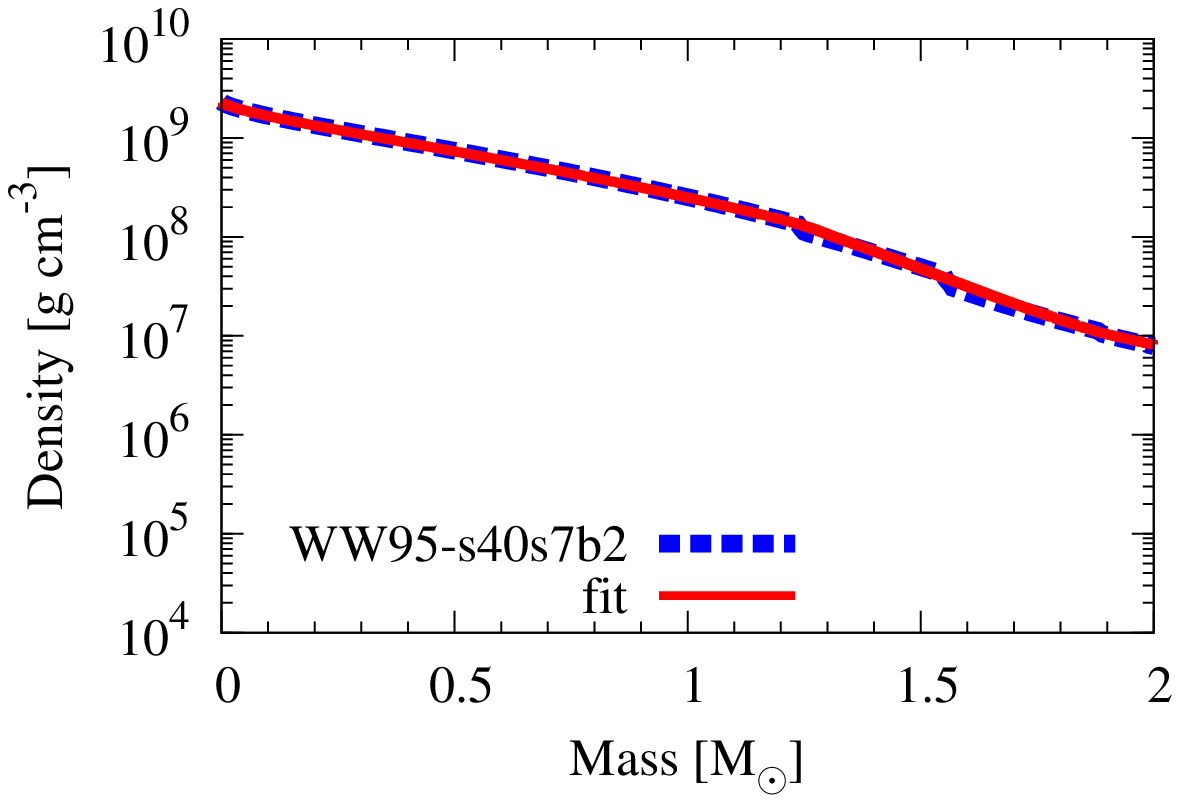}
\includegraphics[width=0.32\textwidth]{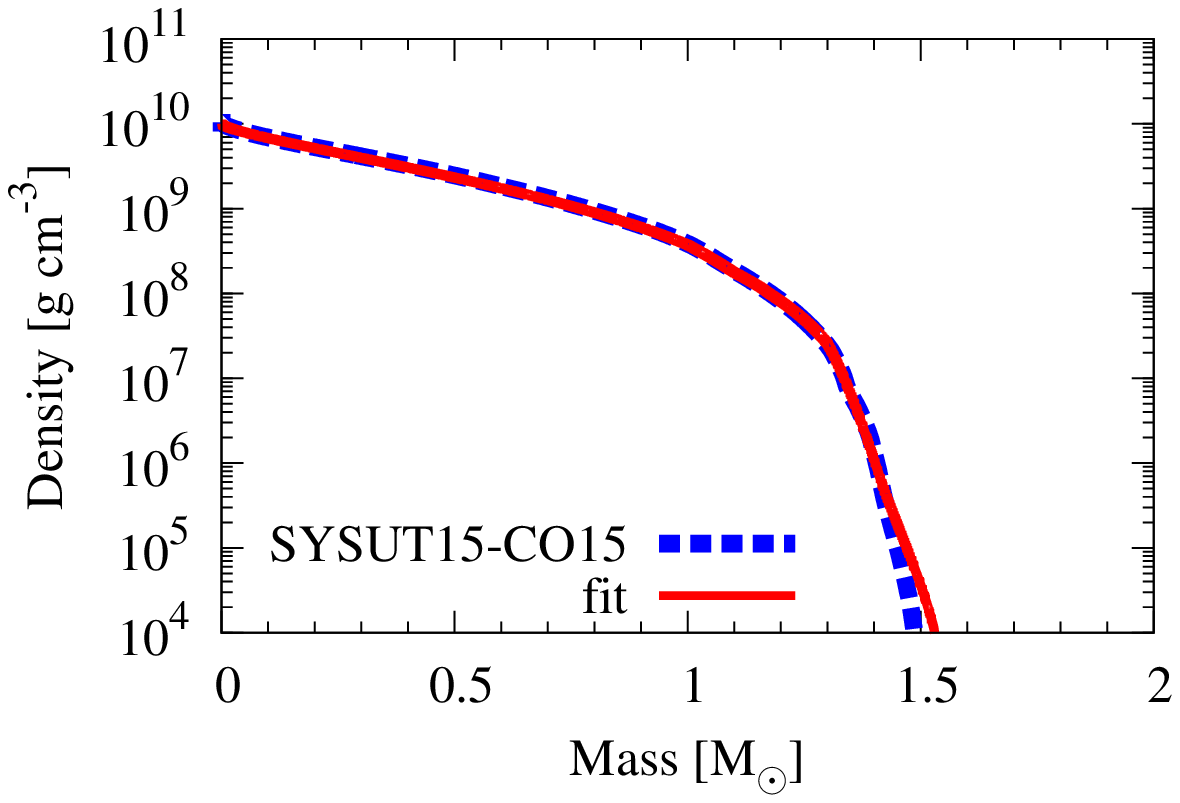}
\caption{Density structures of various stellar evolutionary progenitor
  models (blue dashed lines) and parameterized models fitting these
  models (red solid lines). The corresponding parameters are given in
  Table\,\ref{tab-other}.}
\label{fig:other_prog}
\end{figure*}

In the main text, we discussed a set of models based on the stellar
evolutionary model s11.2 of \cite{woos02}, since it is a well studied
model in the literature. In this appendix, we give the fitting
parameters (Table\,\ref{tab-other}) and density structures
(Fig.\,\ref{fig:other_prog}) of parameterized models reproducing other
typical progenitor models for convenience.

\begin{table*}
\centering
\caption{Parameters for a set of stellar evolutionary models}
\begin{tabular}{lcccccccccccccc}
\hline
Model & $M_1$ & $M_2$ & $M_3$ & $M_4$ & $M_5$ & $S_c$ & $S_1$ & $S_2$ & $S_5$ & $Y_{ec}$ & $Y_{e3}$ & $Y_{e4}$ & $\rho_c$ & $g_\mathrm{eff}$\\
& \multicolumn{5}{c|}{[$M_\odot$]} & \multicolumn{4}{c|}{[$k_B/$baryon]} & & & & [$10^{10}$g cm$^{-3}$] \\
\hline
\multicolumn{14}{c}{Parameterized models fitting models s15.0 and s40.0 of \cite{woos02}}\\
\hline
WHW02-s15.0 & 1.05 & 1.3 & 1.3 & 1.6 & 1.84 & 0.75 & 1.5 & 2.7  & 4.5 & 0.437 & 0.472 & 0.5 & 0.63 & 0.975\\ 
WHW02-s40.0 & 0.98 & 1.55 & 1.55 & 1.84 & 1.86 & 0.96 & 1.6 & 2.6 & 5.8 & 0.443 & 0.48 & 0.5 & 0.37 & 0.975\\
\hline
\hline
\multicolumn{14}{c}{Parameterized model fitting model 8.8$M_\odot$ (O-Ne-Mg core) of \cite{nomo84,nomo87}}\\
\hline
N87-ONeMg & --- & --- & 0.63 & 0.72 & --- & 0.55 & 0.55 & 0.55 & 0.55 & 0.488 & 0.488 & 0.5 & 3.0 & 0.99\\
\hline
\hline
\multicolumn{14}{c}{Parameterized models fitting models N13 and N15 of \cite{nomo88}}\\
\hline
NH88-N13 & 0.68 & 1.11 & 1.15 & 1.17 & 1.18 & 0.65 & 0.97 & 1.54 & 4.2 & 0.406 & 0.462 & 0.5 & 3.0 & 0.975\\
NH88-N15 & 0.7 & 1.2 & 1.3 & 1.31 & 1.38 & 0.74 & 1.01 & 2.26 & 5.0 & 0.411 & 0.472 & 0.5 & 3.1 & 0.975\\
\hline
\hline
\multicolumn{14}{c}{Parameterized models fitting models s12, s15, s20, and s25 of \cite{woos07}}\\
\hline
WH07-s12 & 0.95 & 1.2 & 1.26 & 1.3 & 1.32 & 0.7 & 1.1 & 2.2 & 3.5 & 0.43 & 0.48 & 0.5 & 1.2 & 0.975 \\
WH07-s15 & 1.0 & 1.3 & 1.34 & 1.35 & 1.42 & 0.78 & 1.4 & 2.2 & 3.7 & 0.436 & 0.48 & 0.5 & 0.72 & 0.975 \\
WH07-s20 & 0.96 & 1.5 & 1.54 & 1.81 & 1.82 & 0.93 & 1.56 & 2.62 & 5.0 & 0.443 & 0.482 & 0.5 & 0.36 & 0.975 \\
WH07-s25 & 0.96 & 1.58 & 1.58 & 1.89 & 1.9 & 0.93 & 1.56 & 2.9 & 5.0 & 0.444 & 0.482 & 0.5 & 0.34 & 0.975 \\
\hline
\hline
\multicolumn{14}{c}{Parameterized model fitting model 15$M_\odot$ of \cite{limo06}}\\
\hline
LC06-m15 & 1.12 & 1.38 & 1.38 & 1.6 & 1.78 & 0.66 & 1.5 & 2.7 & 5.0 & 0.454 & 0.48 & 0.5 & 0.52 & 0.975 \\
\hline
\hline
\multicolumn{14}{c}{Parameterized models fitting models s15s7b2, s25s7b8, and s40s7b2 of \cite{woos95}}\\
\hline
WW95-s15s7b2 & 0.95 & 1.28 & 1.28 & 1.42 & 1.43 & 0.7 & 1.3 & 2.1 & 4.5 & 0.432 & 0.476 & 0.5 & 1.0 & 0.975 \\
WW95-s25s7b8 & 1.07 & 1.35 & 1.72 & 2.05 & 2.06 & 1.0 & 1.7 & 2.75 & 5.5 & 0.448 & 0.484 & 0.5 & 0.26 & 0.975 \\
WW95-s40s7b2 & 1.24 & 1.88 & 1.88 & 3.2 & 3.7 & 1.15 & 2.0 & 4.24 & 6.6 & 0.448 & 0.49 & 0.5 & 0.23 & 0.975 \\
\hline
\hline
\multicolumn{14}{c}{Parameterized model fitting model CO15 of \cite{suwa15b}}\\
\hline
SYSUT15-CO15 & 1.0 & 1.1 & 1.11 & 1.3 & 1.42 & 0.65 & 1.25 & 1.9 & 4.7 & 0.43 & 0.468 & 0.5 & 1.0 & 0.99 \\
\hline
\hline
\end{tabular}
\label{tab-other}
\end{table*}

\label{lastpage}

\end{document}